\documentclass[aps,twocolumn,prd,showpacs,nofootinbib]{revtex4}

\usepackage{amsmath}
\usepackage{graphicx}
\usepackage{subfigure}
\usepackage{bm}
\usepackage{amssymb}
\usepackage{latexsym}

\newcommand{\sss}[1]{{\scriptscriptstyle{#1}}}

\newcommand{\lta}{\lesssim}
\newcommand{\gta}{\gtrsim}
\newcommand{\uPl}{\mathrm{Pl}}
\newcommand{\uin}{\mathrm{in}}
\newcommand{\uend}{\mathrm{end}}

\newcommand{\uc}{\mathrm{c}}

\newcommand{\usssPl}{\sss{\uPl}}

\newcommand{\ie}{\textsl{i.e.~}}

\newcommand{\eg}{\textsl{e.g.~}}

\def\spose#1{\hbox to 0pt{#1\hss}}
\def\lta{\mathrel{\spose{\lower 3pt\hbox{$\mathchar"218$}}
     \raise 2.0pt\hbox{$\mathchar"13C$}}}
\def\gta{\mathrel{\spose{\lower 3pt\hbox{$\mathchar"218$}}
     \raise 2.0pt\hbox{$\mathchar"13E$}}}

\newcommand{\de}[2]{\kern - #1 em \mathrm{d} #2}
\newcommand{\ini}{\mathrm{in}}
\newcommand{\cl}{\mathrm{cl}}

\newcommand{\Mp}{M_\usssPl}

\begin{document}

\title{Stochastic Effects in Hybrid Inflation}

\author{J\'er\^ome Martin} \email{jmartin@iap.fr}
\affiliation{Institut d'Astrophysique de Paris, \\ UMR 7095-CNRS,
Universit\'e Pierre et Marie Curie, \\ 98bis boulevard Arago, 75014
Paris, France}

\author{Vincent Vennin} \email{vennin@iap.fr}
\affiliation{Institut d'Astrophysique de Paris, \\ UMR 7095-CNRS,
Universit\'e Pierre et Marie Curie, \\ 98bis boulevard Arago, 75014
Paris, France}

\date{\today}

\begin{abstract}
  Hybrid inflation is a two-field model where inflation ends due to an
  instability. In the neighborhood of the instability point, the
  potential is very flat and the quantum fluctuations dominate over
  the classical motion of the inflaton and waterfall fields. In this
  article, we study this regime in the framework of stochastic
  inflation. We numerically solve the two coupled Langevin equations
  controlling the evolution of the fields and compute the probability
  distributions of the total number of e-folds and of the inflation
  exit point. Then, we discuss the physical consequences of our
  results, in particular the question of how the quantum diffusion can
  affect the observable predictions of hybrid inflation.
\end{abstract}

\maketitle

\section{Introduction}
\label{sec:intro}

Inflation is the leading scenario among the models attempting to
describe the physical conditions that prevailed in the very early
Universe. It consists in a phase of accelerated expansion which
naturally solves the problems of the hot big bang
theory~\cite{Starobinsky:1980te,Guth:1980zm,Linde:1981mu,Albrecht:1982wi,Linde:1983gd}
(for reviews, see
Refs.~\cite{Martin:2003bt,Martin:2004um,Martin:2007bw}). In addition,
it predicts an almost scale invariant power spectrum for the
primordial cosmological fluctuations, the tiny deviations from scale
invariance being related to the microphysics of
inflation~\cite{Mukhanov:1981xt,Mukhanov:1982nu,Hawking:1982cz,Starobinsky:1982ee,Guth:1982ec,Bardeen:1983qw}. As
is well known, this prediction turns out to be fully consistent with
different types of astrophysical observationsm among which is the
measurement of the cosmic microwave background radiation (CMBR)
anisotropies~\cite{Martin:2006rs,Lorenz:2007ze,Lorenz:2008je,Martin:2010kz,Martin:2010hh}.

\par

Inflation is usually driven by one or many scalar fields. In the
context of general relativity, this represents the simplest
mechanism to obtain the negative pressure necessary to produce an
accelerated expansion. However, the physical nature of those scalar
fields is presently unknown, and many different inflationary models
have been suggested. The reason for such a situation originates from
the fact that inflation is a high energy phenomenon. Indeed, the
energy scale of inflation is somewhere between the $\mbox{TeV}$ scale
and the grand unified theory scale~\cite{Martin:2006rs}. At
those scales, particle physics remains elusive, and for a
consequence, there is presently a large variety of different
inflationary scenarios.

\par

However, given the extensions of the standard model of particle
physics, notably those based on supersymmetry and supergravity, it is
clear that some models appear to be more motivated and more generic
than others. In particular, this is the case of hybrid inflation
\cite{Linde:1993cn,Copeland:1994vg}, which can be realized in various
ways in the context of supersymmetry, see for instance the scenarios
named $F$-term inflation and $D$-term inflation (among
others)~\cite{Halyo:1996pp,Binetruy:1996xj,Dvali:1994ms,Kallosh:2003ux}. Hybrid
inflation is a two-field model such that inflation occurs along a
valley in the field space and ends by tachyonic instability along the
so-called waterfall field direction. Hybrid inflation is known to lead
to a blue spectrum for the fluctuations, a prediction which appears to
be disfavored by the most recent
observations~\cite{Martin:2006rs}. However, it was shown
recently~\cite{Clesse:2008pf,Clesse:2010iz,Abolhasani:2010kn} that, in
some regions of the parameter space, a significant number of e-folds
can occur in the waterfall regime. In this case, it was also
demonstrated that the spectral index becomes red, which therefore
implies that the model is in fact totally compatible with the
data~\cite{Clesse:2008pf,Clesse:2010iz}.

\par

In the context of inflation, another interesting question is the role
played by the quantum
corrections~\cite{Vilenkin:1983xp,Starobinsky:1986fx,Goncharov:1987ir,
  Nambu:1987ef,Nambu:1988je,Kandrup:1988sc,Nakao:1988yi,Nambu:1989uf,
  Mollerach:1990zf,Linde:1993xx,Starobinsky:1994bd}. Various works
have shown that they can have a crucial impact on the inflationary
dynamics. This is, for instance, the case for large field inflation if
one starts inflation high enough in the potential. In this case, the
quantum kicks undergone by the field can be so important that the
field climbs its potential instead of rolling down it as should be the
case according to the classical equations of motion. In such a
situation, it is likely that one enters into a regime of eternal
inflation~\cite{Linde:1986fd,Linde:1986fc,Goncharov:1987ir}.

\par

Hybrid inflation is also a model where one expects the quantum
corrections to be very important. It should be the case high in the
inflationary valley but also around the critical point where the
tachyonic instability is
triggered~\cite{GarciaBellido:1996qt,Clesse:2008pf,Clesse:2010iz}. The
goal of this article is to investigate this last question in
detail. In particular, we are interested in whether the quantum
effects can significantly modify the classical dynamics and affect the
observational predictions of the model.

\par

In order to carry out our study, we use the stochastic inflation
formalism~\cite{Starobinsky:1986fx,Nambu:1987ef,Nambu:1988je,
  Kandrup:1988sc,Nakao:1988yi,Nambu:1989uf,Mollerach:1990zf,
  Linde:1993xx,Starobinsky:1994bd,Martin:2005ir,Martin:2005hb}. It
consists in modeling the quantum effects by a stochastic white
noise. As a consequence, the equation describing the motion of the
fields becomes a Langevin equation. As mentioned previously, hybrid
inflation is a genuine two-field model, which implies that one has to
deal with two coupled Langevin equations. Solving this system is a
very complicated task, even if a perturbative expansion is used, as
usually done in the context of single field inflation. This is the
reason why, in this article, we use a numerical approach. This allows
us to compute various interesting quantities such as the probability
density function for the number of e-folds or for the location in
field space of the end of inflation.

\par 

This article is organized as follows. In the next section,
Sec.~\ref{sec:class}, we review in some detail the classical behavior
of the inflaton and waterfall fields. This allows us to clearly
identify the region in the parameter space where a significant number
of e-folds can occur during the waterfall regime. This also permits a
comparison between the classical and stochastic dynamics. In
Sec.~\ref{sec:stocha} we numerically solve the two coupled Langevin
equations that control the behavior of the two fields. We then use
this result to compute various probability density functions, in
particular, that of the number of e-folds and of the inflation exit
point. Finally, in Sec.~\ref{sec:conclusion} we summarize our main
results and present our conclusions.

\section{Classical regimes}
\label{sec:class}

\begin{figure}
\begin{center}
\includegraphics[width=9cm]{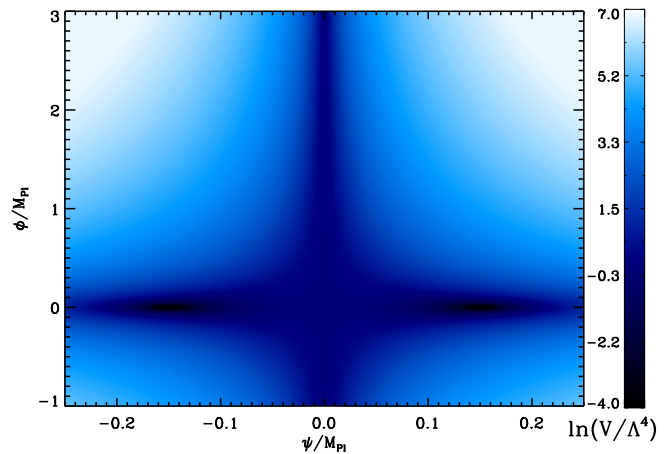}
\caption{Potential of hybrid inflation in the $\left(\phi,\psi\right)$
  plane. The values of the parameters are $\mu=3190.4\, \Mp$,
  $M=\phi_\uc=0.1503\, \Mp$, with $\Mp$ being the reduced Planck mass.}
\label{fig:potential}
\end{center}
\end{figure}

There exist many ways to realize hybrid inflation. In this article,
for simplicity, we focus on the first version studied in
Ref.~\cite{Linde:1993cn}; see also Ref.~\cite{Green:2002wk}. In this
case, the potential in the field space $\left(\phi,\psi\right)$, where
$\phi$ is the inflaton and $\psi$ the waterfall field, is given by the
following expression:
\begin{equation}
V\left(\phi,\psi\right)=\Lambda^4\left[
\left(1-\frac{\psi^2}{M^2}\right)^2+\frac{\phi^2}{\mu^2}
+2\frac{\phi^2\psi^2}{\phi_\mathrm{c}^2M^2}\right].
\end{equation}
It contains four parameters (of dimension one), $\Lambda$, $M$, $\mu$,
and $\phi_\mathrm{c}$. The scale $\Lambda$ is fixed by the Cosmic
Background Explorer (COBE) normalization (the other parameters being
fixed). The true minimums of the potential are located at $\phi=0$ and
$\psi=\pm M$, while the instability point is given by
$\phi=\phi_\mathrm{c}$, $\psi=0$. Along the inflationary valley,
$\psi=0$, the potential reduces to
$\Lambda^4\left[1+(\phi/\mu)^2\right]$ which shows that, in this
regime, inflation cannot end by violation of the slow-roll
conditions. The full hybrid inflation potential is shown in
Fig.~\ref{fig:potential} where the inflationary valley is clearly
visible.

\par

In this section we study the classical behaviors of the inflaton and
waterfall fields. The slow-roll equations controlling the evolution of
the fields can be expressed as
\begin{eqnarray}
\label{KGphi}
3H^2\frac{\mathrm{d}\phi}{\mathrm{d}N} &=& 
-\frac{2\Lambda ^4\phi}{\mu^2}\left(1+\frac{2\psi^2\mu^2}
{\phi_\mathrm{c}^2M^2}\right)\, ,\\
\label{KGpsi}
3H^2\frac{\mathrm{d}\psi}{\mathrm{d}N} &=& -\frac{4\Lambda^4}{M^2}
\psi\left(\frac{\phi^2-\phi_\mathrm{c}^2}{\phi_\mathrm{c}^2}
+\frac{\psi^2}{M^2}\right)\, ,
\end{eqnarray}
where $H=\dot{a}/a$ is the Hubble parameter, $a(t)$ being the
Friedman-Lemaitre-Robertson-Walker (FLRW) scale factor and a dot
denoting a derivative with respect to cosmic time. The quantity $N$ is
the number of e-folds, $N\equiv \ln(a/a_\mathrm{i})$, where
$a_\mathrm{i}$ is the scale factor at the beginning of inflation.

\par

\begin{figure*}
\begin{center}
\includegraphics[width=0.45\textwidth,clip=true]{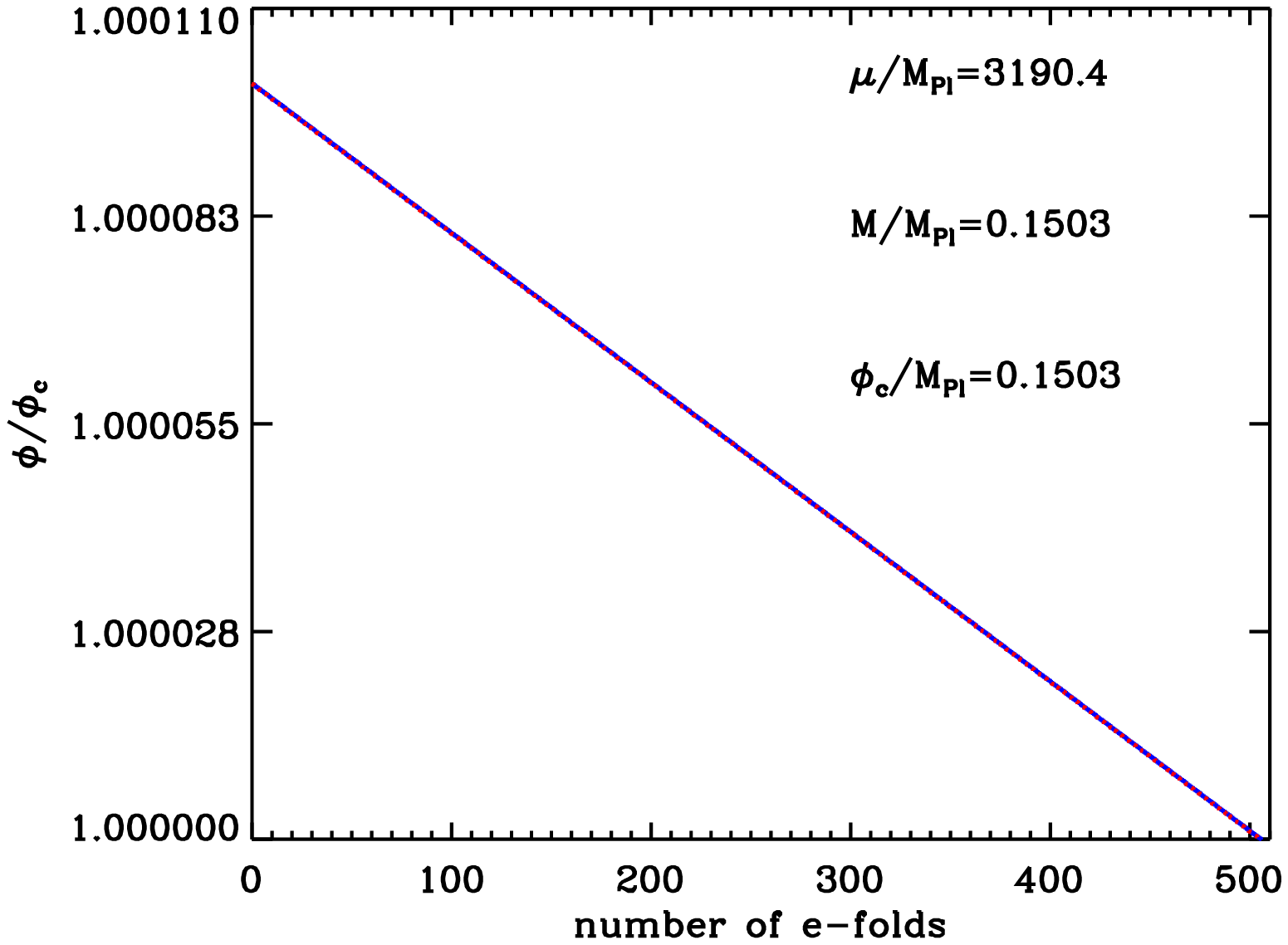}
\includegraphics[height=5.85cm,clip=true]{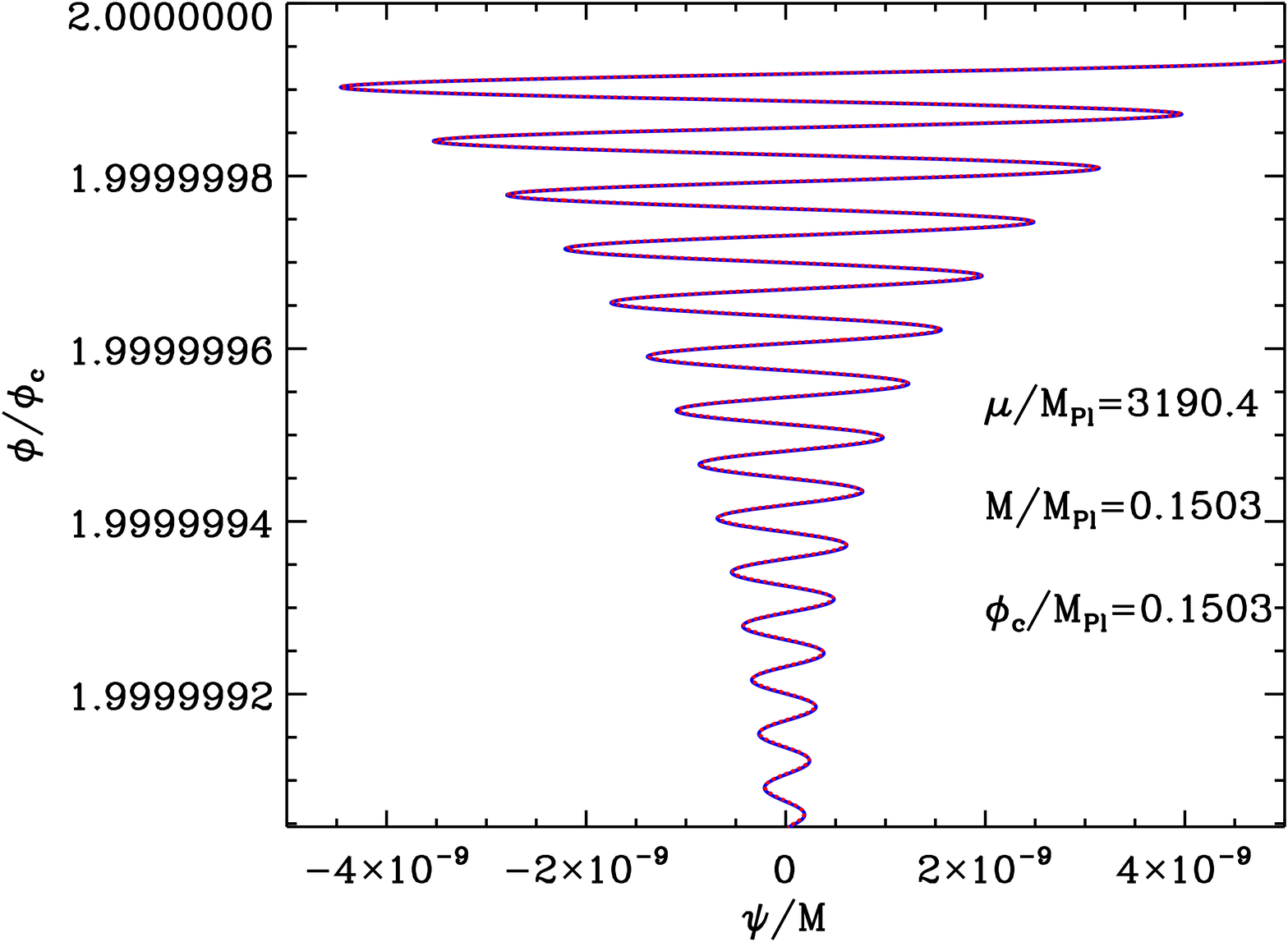}
\caption{Exact numerical solution for the inflaton (left panel, solid
  blue line) and waterfall (right panel, solid blue line) fields in
  the inflationary valley . The red dotted lines represent the
  analytical solution and it is obvious that the approximation is very
  good. The damped oscillations of the waterfall field along the
  inflationary valley are clearly visible. The WKB analytical formula
  given by Eq.~(\ref{psitrajKG2}) is a very good approximation to the
  exact numerical solution.}
\label{fig:valley}
\end{center}
\end{figure*}

In order to study the classical dynamics, it is interesting to
calculate the slow-roll parameters. The hierarchy defined from the
potential~\cite{Stewart:1993bc,Liddle:1994dx,GrootNibbelink:2001qt,Easther:2005nh} is given by the
following expressions:
\begin{eqnarray}
\epsilon_{\phi} &=& \frac{2\phi^2\Mp^2}{\mu^4}\left(1+\frac{2\psi^2\mu^2}
{\phi_\mathrm{c}^2M^2}\right)\, ,\label{epsphi}\\
\epsilon_{\psi} &=& \frac{8\Mp^2\psi^2}{M^4}\left(
\frac{\phi^2-\phi_\mathrm{c}^2}{\phi_\mathrm{c}^2}
+\frac{\psi^2}{M^2}\right)\, ,\label{epspsi}\\
\eta_{\phi \phi} &=& \frac{2\Mp^2}{\mu^2}
\left(1+\frac{2\mu^2\psi^2}{\phi_\mathrm{c}^2M^2}\right)
\, ,\\
\eta_{\phi \psi} &=& \frac{8 \Mp^2 \phi \psi}{\phi_\mathrm{c}^2M^2}\, ,\\
\eta _{\psi \psi} &=& \frac{4\Mp^2}{M^2}\left(\frac{\phi^2-\phi_\mathrm{c}^2}
{\phi_\mathrm{c}^2}+3\frac{\psi^2}{M^2}\right)\, .
\end{eqnarray}
On the other hand, the hierarchy defined from the Hubble parameter,
the so-called Hubble flow parameters
\cite{Schwarz:2001vv,Leach:2002ar}, can be expressed as
\begin{equation}
\epsilon_{n+1}\equiv \frac{{\rm d}\ln \vert \epsilon_n\vert}{{\rm d}N},
\end{equation}
where $\epsilon_0=H_{\rm i}/H(N)$. The above expression implies that
having inflation is strictly equivalent to $\epsilon_1<1$, where
$\epsilon_1=-\dot{H}/H^2$. Obviously, the two hierarchies are related
by simple expressions. In particular, the first horizon flow parameter
is
\begin{equation}
\epsilon_1\simeq\epsilon_\phi+\epsilon_\psi,
\end{equation}
where $\epsilon_{\phi}$ and $\epsilon_{\psi}$ have been defined
before.

\par

Having specified the notation, we now turn to the choice of the free
parameters controlling the shape of the inflationary potential.

\subsection{Physical Priors}
\label{subsec:priors}

It is usually assumed that hybrid inflation occurs in the vacuum
dominated regime, for which $\phi\ll\mu$ and $\psi\ll M$. In this
paper we also assume that this is the case. In any case, otherwise,
hybrid inflation in the valley would be equivalent to a large field
model, which is not the regime of interest here. For simplicity, in
order to reduce the number of free parameters, and as also motivated
by the supersymmetric version of the model, we take $\phi_\uc\simeq
M$. Notice that this assumption does not imply a loss of generality, as
we could easily relax it without drastically modifying the results
obtained in this paper. Finally, in order for inflation to proceed for
small values of the fields (compared to the Planck mass), one can
consider that $\phi_\mathrm{c},M\ll \Mp$, $\Mp$ being the reduced
Planck mass.

\par

In the valley and in the $\phi/\mu\ll 1$ limit, the slow-roll
parameters $\epsilon_1$ and $\epsilon_2$ read
\begin{eqnarray}
\epsilon_1\left(\phi,\psi=0\right)&\simeq &2\frac{\Mp^2}{\mu^2}
\frac{\phi^2}{\mu^2} \, , \\
\epsilon_2\left(\phi,\psi=0\right)&\simeq &4\frac{\Mp^2}{\mu^2}\, .
\end{eqnarray}
Therefore, for the slow-roll approximation to be satisfied, these two
parameters have to be much smaller than $1$, which implies that
$\mu\gg\Mp$.

\par

In the next subsection we study the behavior of the two fields
during the first phase of evolution, namely when the inflaton field is
moving along the valley.

\subsection{The Inflationary Valley}
\label{subsec:valley}

The question of the initial conditions in hybrid inflation is a very
interesting and non-trivial question. It has been studied in detail
in
Refs.~\cite{Clesse:2009zd,Clesse:2010ht,Clesse:2009ur,Clesse:2008pf}.
Here, we simply argue that starting in the valley can be reasonably
justified even if more complicated regimes can be found; see
Ref.~\cite{Clesse:2010ht}. Indeed, if inflation starts beyond the
critical line $\phi=\phi_\uc$, the system very quickly reaches the
region where $\psi/M\ll 1$. In this regime, where the inflaton field
is driving inflation, the slow-roll equation of motion for $\phi$ can
be integrated, leading to
\begin{equation}
N=\frac{1}{4}\frac{\mu^2}{\Mp^2}
\left[\frac{\phi_\mathrm{in}^2}{\mu^2}
-\frac{\phi^2}{\mu^2}
-2\ln\left(\frac{\phi}{\phi_\mathrm{in}}\right)\right]\ ,
\end{equation}
where $\phi_\uin$ denotes the initial value of the inflaton
field. This relation can be inverted, and one
obtains~\cite{Martin:2006rs}
\begin{equation}
\label{phitraj}
\frac{\phi}{\mu}=\left[W_0\left(\frac{\phi_\mathrm{in}^2}{\mu^2}
\mathrm{e}^{\phi_\mathrm{in}^2/\mu^2-4\Mp^2N/\mu^2}\right)\right]^{1/2}\ ,
\end{equation}
where $W_0$ denotes the $0$-branch of the Lambert function. In the
$\phi/\mu\ll 1$ limit, this formula simply reads
\begin{equation}
\label{phitrajphiLTmu}
\phi=\phi_\mathrm{in}\exp\left(-2\frac{\Mp^2}{\mu^2}
N\right)\ .
\end{equation}
This last expression is compared with an exact numerical integration
of the full equations of motion in Fig.~\ref{fig:valley} (left panel),
where it is shown that this is indeed an excellent
approximation. Moreover, this allows us to calculate the number of
e-folds ``generated'' in the valley, which reads
\begin{equation}
\label{Nc}
N_\uc=\frac{\mu^2}{2\Mp^2}\ln\left(\frac{\phi_\uin}{\phi_\uc}\right)\, .
\end{equation} 
Clearly, this number is large because $\mu\gg \Mp$.

\par

\begin{figure}
\includegraphics[width=8.5cm]{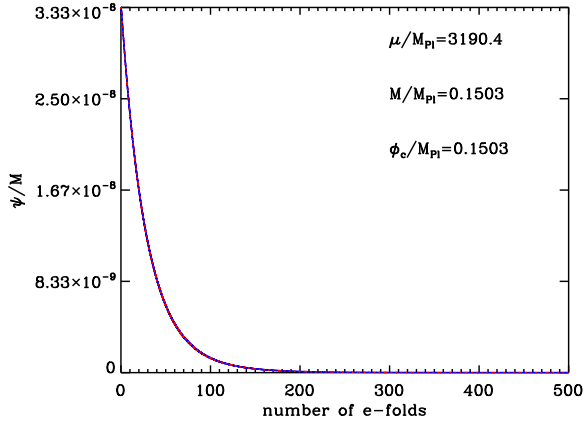}
\caption{Exact numerical solution for the waterfall field (solid blue
  line) in the inflationary valley after the oscillatory regime. The
  red dotted line represents the analytical solution, and it is obvious
  that the approximation is very good.}
\label{fig:damped}
\end{figure}

Let us now study the behavior of the waterfall field in the vicinity
of the valley, when $\psi/M\ll 1$. Since $\psi$ undergoes damped
oscillations in this regime, it is clear that the slow-roll
approximation cannot be used. On the other hand, since $\psi $
oscillates much faster than $\phi $ moves, the
Wentzel-Kramers-Brillouin (WKB) approximation can be used to describe
this regime. The solution can be expressed as
\begin{eqnarray}
\label{psitrajKG2}
\psi&=&\psi_\mathrm{in}
\frac{{\rm e}^{-3N/2}}{\sqrt{2\omega (N)}}
\left[C_1{\rm e}^{I\left(N\right)}
+C_2{\rm e}^{-I\left(N\right)}\right],
\end{eqnarray} 
where $\psi_\uin$ is the initial value of the waterfall field, and where
$I\left(N\right)$ is defined by the following expression (an
unimportant sign has been ignored):
\begin{eqnarray}
  I\left(N\right)&\equiv & i\int \omega(n){\rm d}n, \\
  &=&\frac{3}{2}\int_{0}^{N}\sqrt{1-\frac{16}{3}\frac{\Mp^2}{M^2}
    \frac{\phi^2/\phi_\mathrm{c}^2-1}{1+\phi^2/\mu^2}}\,\mathrm{d}n,
\end{eqnarray}
and where $C_1$ and $C_2$ are integration constants. The validity of
the WKB approximation can be checked by estimating the following
quantity, $({\rm d}\omega/{\rm d}N) /\omega \simeq {\cal
  O}(1)\Mp^2/\mu^2\ll 1$. From the above expression one notices that
oscillations in the $\psi$ direction occur in the regime
\begin{equation}
\frac{16}{3}\frac{\Mp^2}{M^2}\frac {\phi^2/
\phi_\mathrm{c}^2-1}{1+\phi^2/\mu^2}>1\, ,
\end{equation} 
that is to say, in the region
\begin{equation}
\label{eq:conditionosci}
\frac{\phi}{\phi_\uc}>1+\frac{3}{32}\frac{M^2}{\Mp^2}.
\end{equation} 
Since $M\ll \Mp$, we see that the field oscillates almost all the time
before the critical point is met. In fact, it turns out that the
integral $I(N)$ can be performed. One finds
\begin{eqnarray}
I(N)&\simeq & -\sqrt{3}\frac{\Mp}{M}\frac{\mu^2}{\Mp^2}
\Biggl[\sqrt{\frac{\phi^2}{\phi_\mathrm{c}^2}-1}
-\arctan \left(\sqrt{\frac{\phi^2}{\phi_\mathrm{c}^2}-1}\right)
\nonumber \\ 
&-&\sqrt{\frac{\phi_\ini^2}{\phi_\mathrm{c}^2}-1}
+\arctan \left(\sqrt{\frac{\phi^2_\ini}{\phi_\mathrm{c}^2}-1}\right)
\Biggr].
\end{eqnarray}
This solution is compared with the exact numerical solution in
Fig.~\ref{fig:valley} (right panel). Clearly, the approximation is
excellent.

\par

When the condition~(\ref{eq:conditionosci}) is not satisfied,
$I(N)\simeq \pm 3N/2+\cdots$ and the oscillations stop. Since the
gradients become small, this time, one can use the slow-roll
approximation in order to describe the motion of $\psi$. Notice that,
since $M$ is small (in Planck units), the above-mentioned regime
occurs for a very small range of values for $\phi$. However, a large
number of e-folds $\propto \mu^2M^2/\Mp^4$ can be produced during this phase.
The slow-roll equation of motion for $\psi$ can be straightforwardly 
integrated and gives
\begin{equation}
\psi=\psi_\mathrm{in}\exp\left[4\frac{\Mp^2}{M^2}
\int_{0}^{N}\frac{1-\phi^2\left(n\right)/\phi_\mathrm{c}^2}
{1+\phi^2\left(n\right)/\mu^2}
\,\mathrm{d}n\right]\ ,
\end{equation}
where $\phi\left(n\right)$ is given by Eq.~(\ref{phitraj}). Since
$\phi>\phi_\mathrm{c}$ in the valley, $\psi$ decreases with $N$ and
obviously remains in the $\psi\ll M$ region. If one uses the fact that $\phi\ll
\mu$, then the integral in the above formula can be performed
exactly. Upon using Eq.~(\ref{phitrajphiLTmu}), one obtains
\begin{eqnarray}
\label{psitrajSRphiLTmu}
\psi &=&\psi_\uin\exp\left[4\frac{\Mp^2}{M^2}N
-\frac{\phi_\uin^2}{\phi_\uc^2}\frac{\mu^2}{M^2}\right.\nonumber\\
& &+\left.\frac{\phi_\uin^2}{\phi_\uc^2}\frac{\mu^2}{M^2}
\exp\left(-4\frac{\Mp^2}{\mu^2}N\right)\right]
\nonumber\\
&\simeq&\psi_\uin\exp\left[-4\frac{\Mp^2}{M^2}
\left(\frac{\phi_\uin}{\phi_\uc}-1\right)N\right]\, ,
\end{eqnarray}
where, in the last equation, we have used the fact that $\mu\gg \Mp$.
In that case, one concludes that $\psi$ is exponentially damped after
the oscillations have stopped and before the critical point is
reached. To our knowledge, this regime was not considered before. The
above expression is compared with an exact numerical integration of
the full equations of motion in Fig.~\ref{fig:damped}. As one can
notice, the agreement between the exact numerical solution and the
analytical approximated expression is excellent. The previous formula
also allows us to calculate the classical value of $\psi$ when the
system reaches the critical point. It is given by
\begin{equation}
\label{psic}
\psi_\uc=\psi_\uin\exp\left[-2\frac{\mu^ 2}{M^ 2}
\left(\frac{\phi_\uin}{\phi_\uc}-1\right)
\ln\left(\frac{\phi_\uin}{\phi_\uc}\right)\right]\, .
\end{equation}
In practice, this value is always extremely small, thanks to the fact
that $\mu\gg \Mp^2$ and, as was noticed in
Refs.~\cite{GarciaBellido:1996qt,Clesse:2010iz}, the quantum
fluctuations of $\psi$ can be much larger than its classical value. We
come back to this point in detail in the next section. However,
before addressing this issue, in the next subsection, we describe the
classical motion of the two fields during the waterfall stage.

\subsection{The Waterfall Regime}
\label{subsec:waterfall}

The waterfall regime has recently been studied by various authors; see
\eg
Refs.~\cite{Abolhasani:2011yp,Lyth:2010ch,Clesse:2010iz,Abolhasani:2010kr,Kodama:2011vs}.
Here, we mainly follow the terminology used in
Ref.~\cite{Kodama:2011vs}. We assume that slow roll is valid
initially, at the critical point. We first study the so-called
``phase $0$''~\cite{Kodama:2011vs}. It consists in neglecting the last
term in the inflaton slow-roll equation~(\ref{KGphi}) and the first
one in the right-hand side of the waterfall equation~(\ref{KGpsi}) (on
the grounds that, initially, $\phi=\phi_\mathrm{c}$). Notice that, in
this case, one could also solve the full inflaton equation, keeping
the second time derivative, since in this approximation it becomes
linear. On the other hand, the waterfall equation is nonlinear. In
this sense, we do not start from a linear situation. It is easy to
find the (slow-roll) solutions, and they read
\begin{eqnarray}
\label{eq:solphi0}
\phi(N) &=& \phi_\mathrm{c} \exp\left[-2\frac{\Mp^2}{\mu^2}
\left(N-N_\mathrm{c}\right)\right]\, ,\\
\label{eq:solpsi0}
\psi(N) &=& \psi _\mathrm{c} \left[1+\frac{8\Mp^2\psi_\mathrm{c}^2}{M^4}
\left(N-N_\mathrm{c}\right)\right]^{-1/2}\, ,
\end{eqnarray} 
where $N_\mathrm{c}$ denotes the number of e-folds at the critical
point, \ie at the onset of the waterfall phase. In field space the
trajectory reads
\begin{equation}
\phi=\phi_\mathrm{c}\exp \left[-\frac{M^4}{4\mu^2\psi_\mathrm{c}^2}
\left(\frac{\psi_\mathrm{c}^2}{\psi^2}-1\right)\right].
\end{equation}
Of course, instead of expressing $\phi $ in terms of $\psi$, one can
also express $\psi $ in terms of the inflaton field. In this case one
obtains
\begin{equation}
\psi =\psi_\mathrm{c}\left[1-\frac{4\mu^2\psi_\mathrm{c}^2}{M^4}
\ln \left(\frac{\phi}{\phi_\mathrm{c}}\right)\right]^{-1/2}.
\end{equation}
These expressions are fully consistent with Ref.~\cite{Kodama:2011vs}.

\par

The next question is when phase $0$ stops. By definition, upon
using Eqs.~(\ref{KGphi}) and~(\ref{KGpsi}), it occurs when
$\phi=\phi_1$ and $\psi=\psi_1$ such that
\begin{equation}
-\frac{\phi^2_1}{\phi_\mathrm{c}^2}+1=\frac{\psi^2_1}{M^2}.
\end{equation}
Indeed, among the two conditions that we have required in order to
derive the solutions~(\ref{eq:solphi0}) and~(\ref{eq:solpsi0}), this
one is the first to be violated since $\psi(N)$ decreases during
phase $0$. This condition can also be written as
\begin{equation}
2\ln \frac{\phi_1}{\phi_\mathrm{c}}=\ln \left(1-\frac{\psi_1^2}{M^2}
\right)\simeq -\frac{\psi^2_1}{M^2}.
\end{equation}
Then, using the slow-roll trajectory one easily finds that
\begin{equation}
\ln \frac{\phi_1}{\phi_\mathrm{c}}\simeq \frac{M^4}{8\mu^2\psi_\mathrm{c}^2}
\left(1-\sqrt{1+\frac{8\mu^2 \psi_\mathrm{c}^4}{M^6}}\right),
\end{equation}
and
\begin{equation}
\psi_1\simeq M\sqrt{-2\ln \frac{\phi_1}{\phi_\mathrm{c}}}\, .
\end{equation}
If we are in the regime where $8\mu^2\psi_\mathrm{c}^4/M^6\ll 1$, then
one has
\begin{eqnarray}
\label{eq:expphi1}
\ln \frac{\phi_1}{\phi_\mathrm{c}} &\simeq & 
-\frac{\psi_\mathrm{c}^2}{2M^2}+\frac{\mu^2\psi_\mathrm{c}^6}{M^8}+\cdots\, ,
\\
\label{eq:exppsi1}
\psi_1&\simeq &\psi_\mathrm{c}\left(1
-\frac{\mu^2\psi_\mathrm{c}^4}{M^6}+\cdots\right).
\end{eqnarray}
From these expressions one can easily estimate the number of e-folds
in phase $0$. One obtains
\begin{equation}
N_1-N_\mathrm{c}\simeq \frac{\mu^2\psi_\mathrm{c}^2}{4\Mp^2M^2}+\cdots \ll 1,
\end{equation}
where $N_1$ denotes the number of e-folds at the end of phase $0$. We
see that the above quantity [as well as the parameter used in the
expansion that leads to Eqs.~(\ref{eq:expphi1})
and~(\ref{eq:exppsi1})] is controlled by $\mu /\Mp$, which is large,
and by $\psi_{\mathrm{c}}/M$ which is small. Therefore, the smallness
of this parameter is a priori not obvious. The two situations, where
it is small or large, have been studied in
Ref.~\cite{Kodama:2011vs}. However, in practice, $\psi_\mathrm{c}/M$
is so small that the parameter mentioned previously is always
small. In this case, we conclude that phase $0$ is unimportant
since it lasts a negligible number of e-folds. As a consequence, the
values of $\phi$ and $\psi$ remain almost unchanged during that phase.

\begin{figure*}
\begin{center}
\includegraphics[width=0.45\textwidth,clip=true]{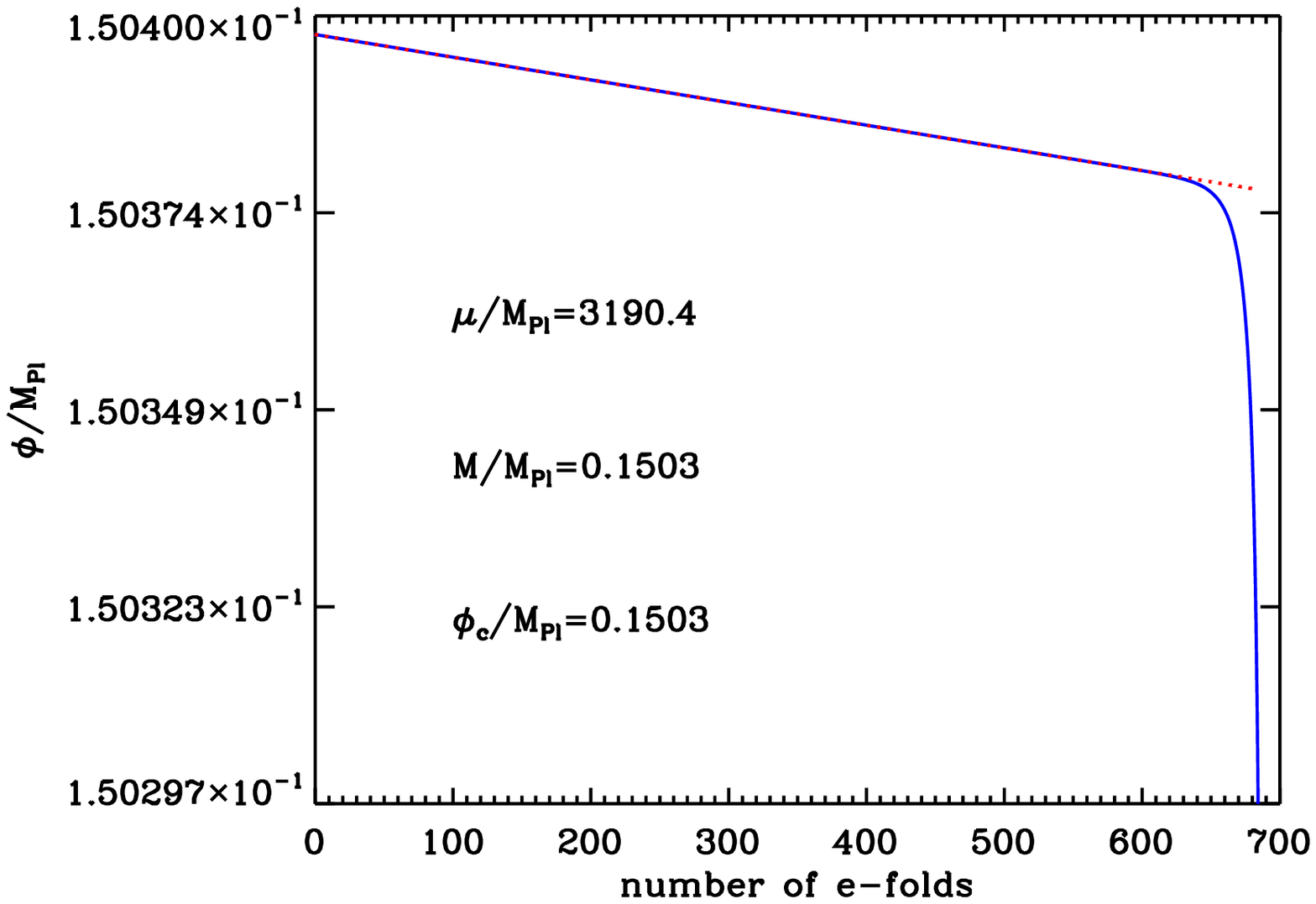}
\includegraphics[width=0.45\textwidth,clip=true]{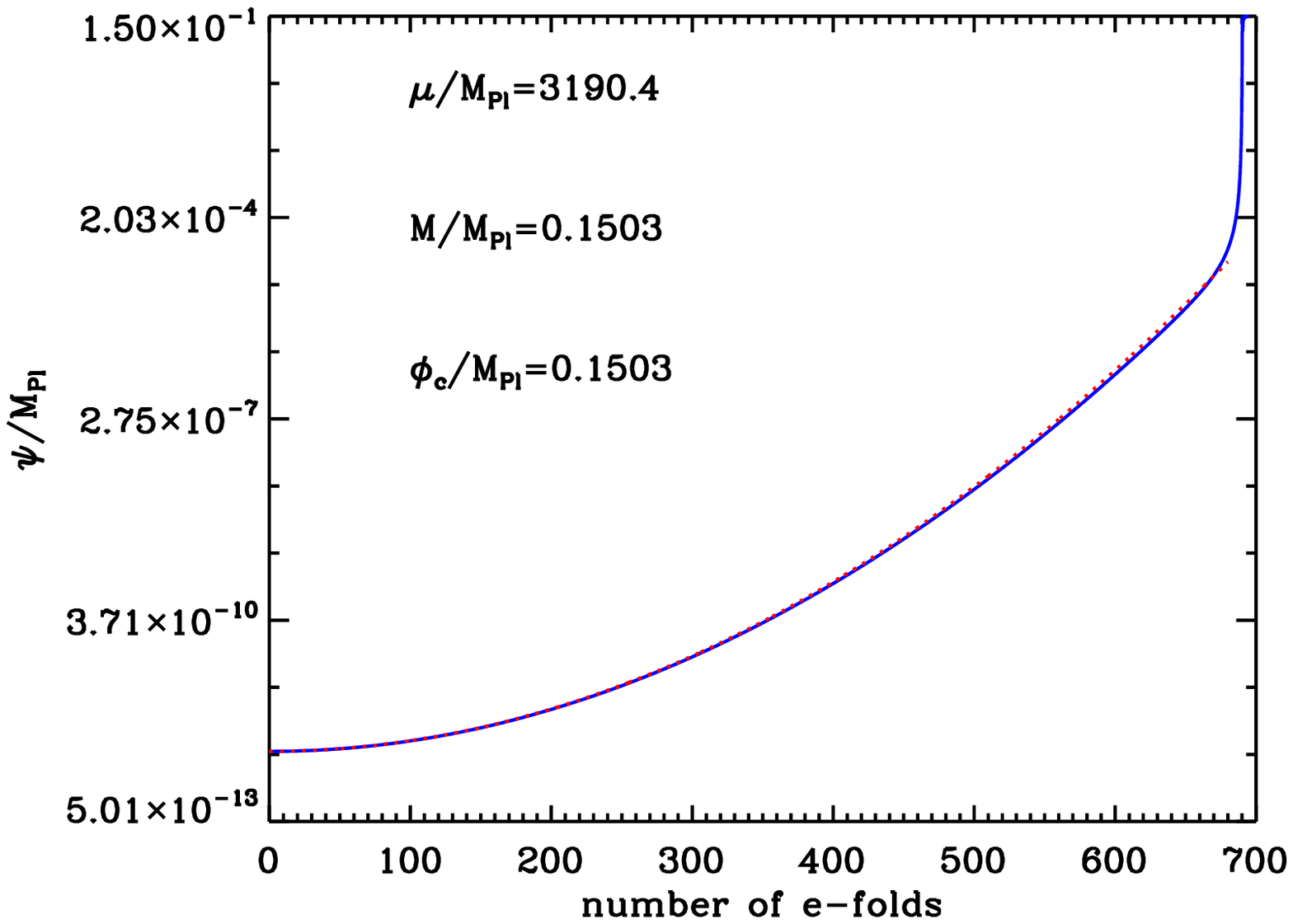}
\caption{Exact numerical solution for the inflaton (left panel, solid
  blue line) and waterfall (right panel, solid blue line) fields. The
  red dotted lines represents the slow-roll analytical solution during
  phase $1$.}
\label{fig:waterfall}
\end{center}
\end{figure*}

We now proceed with phase $1$. By definition, the second term on the
right-hand side of the waterfall equation~(\ref{KGpsi}) can be
neglected. This means that this equation, as was already the case
for the inflaton equation of motion (which remains unaffected during
phase $1$), becomes linear. For this reason, sometimes, this phase is
also called the ``linear phase.'' During this phase, the solution for
the inflaton field is unchanged but, of course, one now has to solve
the new approximated equation for the waterfall field . The solution
can be easily calculated and reads
\begin{eqnarray}
\ln \frac{\psi}{\psi_1}&=&\frac{\mu^2}{M^2}
\left[{\rm e}^{-4\Mp^2(N-N_\mathrm{c})/\mu^2}-{\rm e}^{-4\Mp^2(N_1-N_\mathrm{c})/\mu^2}
\right]\nonumber \\ & &+\frac{4\Mp^2}{M^2}\left(N-N_1\right).
\end{eqnarray}
Then, one can Taylor expand the exponential functions since we are in
the regime where $\mu/\Mp\gg 1$. This gives
\begin{equation}
\psi=\psi_1\exp\left\{\frac{8\Mp^4}{\mu^2M^2}
\left[\left(N-N_\mathrm{c}\right)^2 -\left(N_1-N_\mathrm{c}\right)^2\right]\right\}.
\end{equation}
This solution together with the solution for the inflaton is
represented in Fig.~\ref{fig:waterfall}. This plot confirms that the
previous approximated solutions match the exact ones with a very good
precision. Finally, in field space, the trajectory now reads
\begin{equation}
\psi=\psi_1\exp\left[\frac{2\mu^2}{M^2}\left(\ln ^2 \frac{\phi}{\phi_\mathrm{c}}
-\ln ^2 \frac{\phi_1}{\phi_\mathrm{c}}\right)\right].
\end{equation}
or, equivalently,
\begin{equation}
\ln ^2\frac{\phi}{\phi_\mathrm{c}}=\ln^2\frac{\phi_1}{\phi_\mathrm{c}}+
\frac{M^2}{2\mu^2}\ln \frac{\psi}{\psi_1}.
\end{equation}

\par

Phase $1$ stops when, on the right-hand side of the slow-roll
inflaton equation of motion~(\ref{KGphi}) for $\phi$, the last term
becomes $1$. This occurs for $\psi\equiv \psi_2$, where
\begin{equation}
\psi_2^2=\frac{\phi_\mathrm{c}^2M^2}{2\mu^2}\, ,
\end{equation}
and $\phi=\phi_2$ with
\begin{eqnarray}
\ln ^2\frac{\phi_2}{\phi_\mathrm{c}}&\simeq & \ln^2\frac{\phi_1}{\phi_\mathrm{c}}+
\frac{M^2}{2\mu^2}\ln \left(\frac{\phi_\mathrm{c}M}{\sqrt{2}\mu \psi_1}\right) \\
&\simeq & \frac{M^2}{2\mu^2}\ln \left(\frac{\phi_\mathrm{c}M}{\sqrt{2}
\mu \psi_\mathrm{c}}\right)\, ,
\end{eqnarray}
the last approximated relation being obtained under the assumption
that phase $0$ can be neglected and, as a consequence, that
$\phi_1\simeq \phi_\mathrm{c}$ and $\psi_1\simeq \psi_\mathrm{c}$. It
is also important to realize that the terms
$1-\phi^2/\phi_\mathrm{c}^2$ and $\psi^2/M^2$ are equal at the onset
of phase $1$ and then both increase. It is therefore necessary to
check that, at the end of phase $1$, the term
$1-\phi^2/\phi_\mathrm{c}^2$ still dominates over $\psi^2/M^2$. In
other words, one has to verify that $\psi^2/M^2$ has increased less
rapidly than $1-\phi^2/\phi_\mathrm{c}^2$. Using the solution for the
waterfall, one has
\begin{equation}
N_2-N_\mathrm{c}\simeq \frac{\mu M}{2\sqrt{2}\Mp^2}\ln ^{1/2}\left(
\frac{\psi_2}{\psi_\mathrm{c}}\right),
\end{equation}
where $N_2$ denotes the number of e-folds at the end of phase $1$
or, equivalently, at onset of the phase $2$. Upon using this
formula, this leads to
\begin{equation}
\frac{\phi_2^2}{\phi_\mathrm{c}^2}-1=-\sqrt{2}\frac{M}{\mu}
\ln ^{1/2}\left(
\frac{\phi_\mathrm{c}M}{\sqrt{2}\mu\psi_\mathrm{c}}\right)\, ,
\end{equation}
an expression that should be compared with
\begin{equation}
\frac{\psi_2^2}{M^2}=\frac{\phi_\mathrm{c}^2}{2\mu^2}.
\end{equation}
We see that the condition $\phi_2^2/\phi_\mathrm{c}^2-1\gg
\psi_2^2/M^2$ is a priori not obvious. However, in the case under
scrutiny in this article, one chooses $\phi_\mathrm{c}$ and $M$ to be
roughly of the same order of magnitude and $\mu\gg \Mp$. As a
consequence, the condition is satisfied since
$\phi_2^2/\phi_\mathrm{c}^2-1$ scales as the inverse of $\mu$
(neglecting the influence of the logarithm) while $\psi_2^2/M^2$
scales as the inverse of $\mu^2$. However, it is also clear that one
could easily design a situation where this is not true. Here, we
restrict ourselves to situations where this does not happen.

\par

Finally, let us express the number of e-folds produced during
phase $1$. It is given by
\begin{equation}
\label{eq:efoldswater}
N_2-N_\mathrm{c}\simeq \frac{\mu M}{2\sqrt{2}\Mp^2}\ln ^{1/2}\left(
\frac{\phi_\mathrm{c}M}{\sqrt{2}\mu \psi_\mathrm{c}}\right).
\end{equation}
Upon using Eq.~(\ref{psic}), one could also replace $\psi_\uc$ by its
expression in the above equation to obtain a formula depending on the
initial conditions. We see that the number of e-folds during
phase $1$ is essentially controlled by the ratio $\mu M/\Mp^2$. This
conclusion is in agreement with the results of Ref.~\cite{Kodama:2011vs}. As a
consequence, for $\mu M/\Mp^2>1$, $N_2-N_\mathrm{c}$ can be
large. Hence, we conclude that the number of e-folds during the
waterfall phase can indeed be much greater than the $60$ required for
inflation to be successful as was noticed in Ref.~\cite{Clesse:2010iz}. The
previous considerations allow us to identify where, in the parameter
space, this regime occurs. We have studied this classical phase of
evolution in some detail because this regime is of particular
interest for the present article. Indeed, in the next section, we show
that in this case the quantum effects play an important role.

\begin{figure}
\begin{center}
\includegraphics[width=8.5cm]{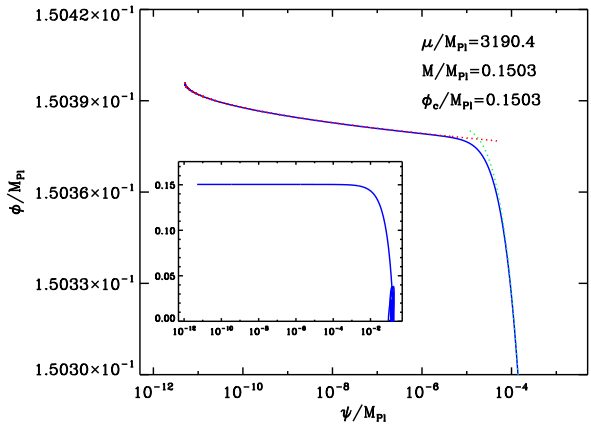}
\caption{Classical background evolution of the inflaton and waterfall
  fields starting from the critical point, $\phi=\phi_\mathrm{c}$ and
  $\psi=10^{-12}\Mp$. The solid blue curve represents the exact (\ie
  numerical) trajectory while the dotted red curve is the slow-roll
  approximation during phase $1$ and the dotted green curve is the
  slow-roll approximation during phase $2$. The inset shows the
  overall evolution of the two fields with, in particular, the
  oscillations around the true minimum of the potential at the end of
  inflation.}
\label{fig:trajec_field}
\end{center}
\end{figure}

\par

Let us now briefly mention phase $2$ (it was studied in more
detail in Ref.~\cite{Kodama:2011vs}). This time one needs to keep the
last term in the inflaton equation of motion~(\ref{KGphi}). This means
that the evolution for $\phi$ is modified and, as a consequence, the
formula giving $\psi(N)$ is no longer valid since it made use of the
evolution for $\phi$ established before. In this regime the
Eq.~(\ref{KGphi}) and~(\ref{KGpsi}) become fully
coupled. However, it is still possible to find the slow-roll
trajectory in the field space. One obtains~\cite{Kodama:2011vs}
\begin{equation}
\frac{{\rm d}\phi}{{\rm d}\psi}=\frac{\phi \psi}{\phi^2-\phi^2_\mathrm{c}}\, 
\end{equation}
which can be easily integrated, and the solution reads
\begin{equation}
\psi^2=\psi_2^2+\phi^2-\phi_2^2-2\phi_\mathrm{c}^2\ln \frac{\phi}{\phi_2}.
\end{equation}
This expression (green dotted line) is compared to the exact numerical
solution (blue solid line) in Fig.~\ref{fig:trajec_field}. Clearly,
the agreement is excellent. During phase $2$, inflation stops and
the system starts oscillating around one of the two true minimums of
the potential. This is the onset of the reheating phase.

\par

The above considerations complete this section. Having mastered the
classical dynamics of the fields in the valley and during the
waterfall regime, we are now in a position where we can turn to the
main topic of this article, namely, studying the role played by the
quantum effects. This is the goal of the next section.

\section{Stochastic Effects}
\label{sec:stocha}

In this article, we use the stochastic inflation formalism to study
the quantum effects. In this formalism, an effective Langevin equation
can be derived for the ``coarse-grained'' field, \ie the original
field averaged over a physical volume the size of which is typically
larger than the Hubble radius $H^{-1}$. Applied to the case of hybrid
inflation, one obtains two coupled Langevin equations for the inflaton
and the waterfall fields, respectively. They read
\begin{eqnarray}
\label{Langphi}
3H^2\frac{\mathrm{d}\phi}{\mathrm{d}N} &=& 
-\frac{2\Lambda ^4\phi}{\mu^2}
\left(1+\frac{2\psi^2\mu^2}{\phi_\mathrm{c}^2M^2}\right)
+\frac{3H^3}{2\pi}\xi_\phi\left(N\right)\, ,\\
\label{Langpsi}
3H^2\frac{\mathrm{d}\psi}{\mathrm{d}N} &=& 
-\frac{4\Lambda^4}{M^2}\psi\left(\frac{\phi^2-\phi_\mathrm{c}^2}
{\phi_\mathrm{c}^2}+\frac{\psi^2}{M^2}\right)
+\frac{3H^3}{2\pi}\xi_\psi\left(N\right)\, ,\nonumber  \\
\end{eqnarray}
where $\xi_\phi$ and $\xi_\psi$ are two uncorrelated white Gaussian
noises with $0$-mean and $1$-variance. Notice that the time variable
used is the number of e-folds. It was argued in
Refs.~\cite{Finelli:2008zg,Finelli:2010sh,Finelli:2011gd} that this
choice is preferred.

\subsection{Can the Quantum Effects Be Important?}
\label{subsec:important?}

\begin{figure*}
\begin{center}
\includegraphics[width=0.45\textwidth,clip=true]{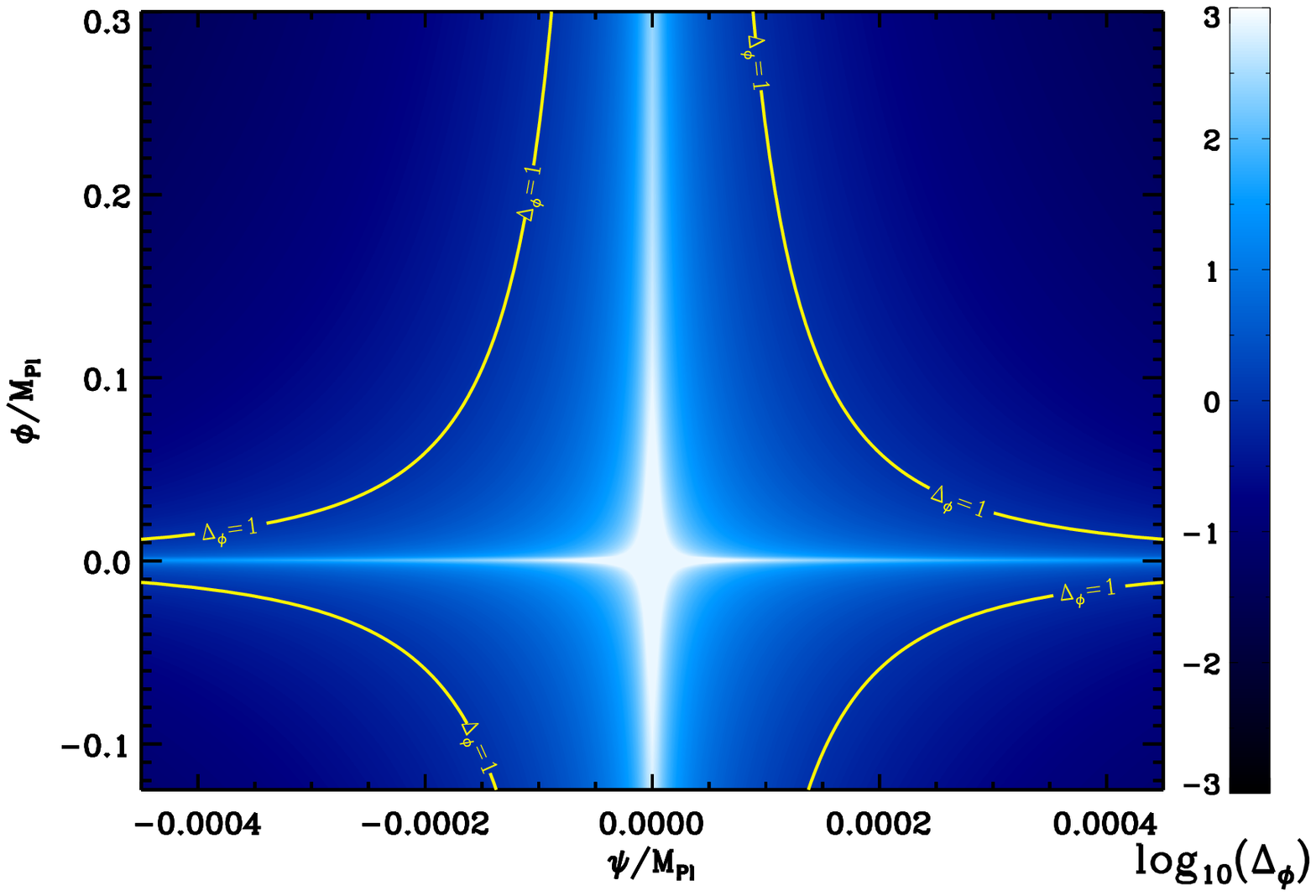}
\includegraphics[width=0.45\textwidth,clip=true]{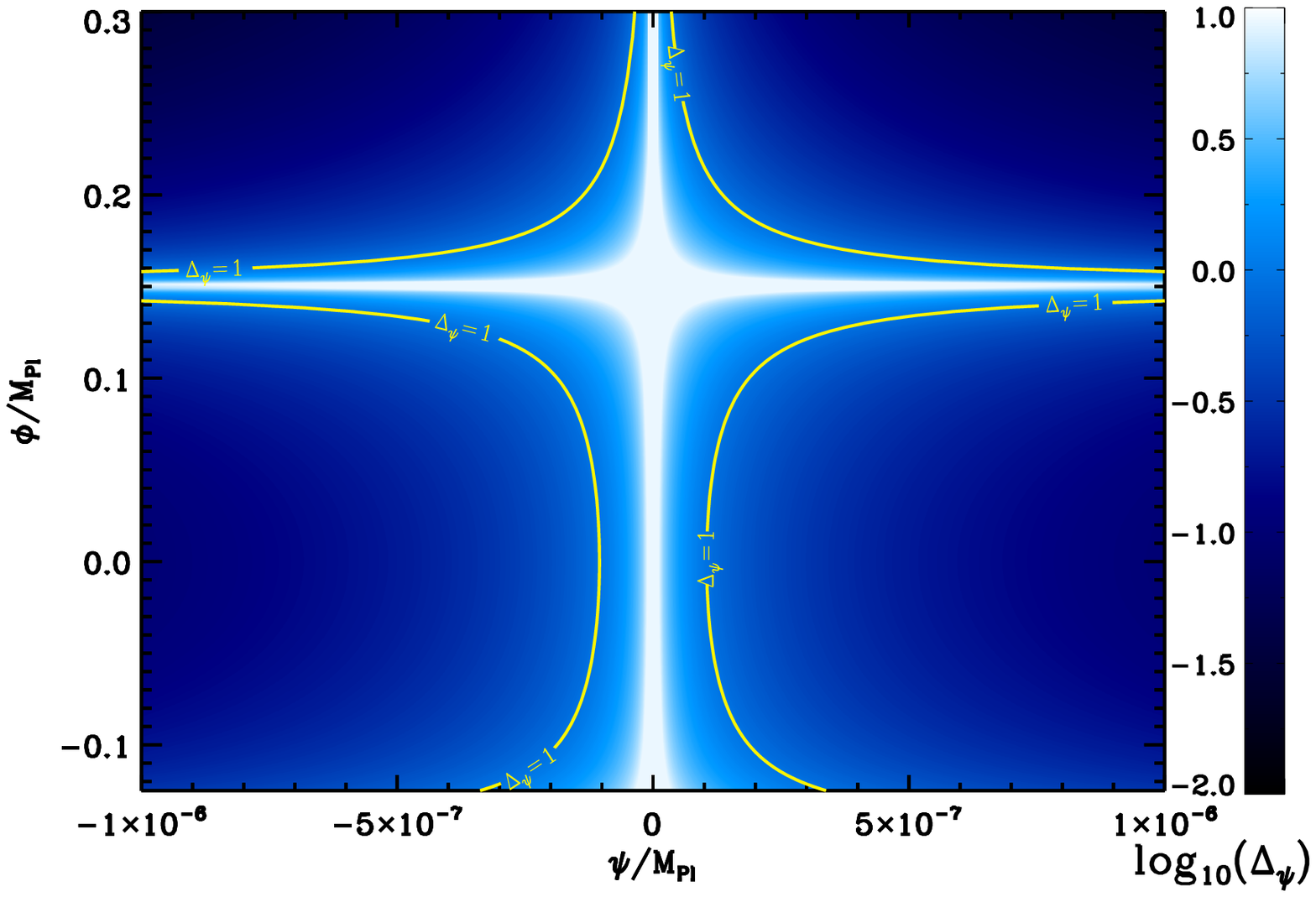}
\caption{$\Delta_\phi$ and $\Delta_\psi$ plotted in the
  $\left(\phi,\psi\right)$ plane for the parameters $\mu=3190.4\,\Mp$,
  $M=\phi_\uc=0.1503\,\Mp$, $\Lambda=0.01418\,\Mp$. $\Delta_\phi$
  and/or $\Delta_\psi$ greater than $1$ ($\sim $ in white on the plot)
  signal that the quantum effects are dominant. The stochastic effects
  in the $\psi$ direction obviously dominate over the classical
  contributions in the valley and around the critical point, while the
  stochastic effects in the $\phi$ direction dominate in the valley
  and around the origin.}
\label{fig:critere}
\end{center}
\end{figure*}

Having at our disposal the two Langevin equations presented above, the
first question is whether the stochastic noises can really play an
important role and, if so, where in the field plane. This issue can be
addressed in the following manner. During a typical time interval
$\Delta t=H^{-1}$, both stochastic and classical evolutions of the
fields $\phi$ and $\psi$ can be read off directly from
Eqs.~(\ref{Langphi}) and (\ref{Langpsi}). Roughly speaking, the
typical classical change in the inflaton value is $\simeq
\Mp^2(\partial V/\partial \phi)/V$, while the magnitude of the quantum
kick is $H/(2\pi)$. Therefore, in order to assess the relative
contribution of the stochastic effects over the classical ones, one
can study the ratios $\Delta_\phi$ and $\Delta_\psi$ of these two
quantities for each field (in the context of a quartic large field
model, this is how one can deduce that the quantum corrections
dominate if the value of the field is larger than $\lambda
^{-1/6}\Mp$, where $\lambda$ is the self-coupling constant that
appears in the potential). This amounts to taking $\Delta_\phi \equiv
V^{3/2}/[2\pi \sqrt{3}\Mp^3(\partial V/\partial \phi)]$ and a similar
definition for $\Delta _\psi$. In the vacuum dominated regime, the two
quantities $\Delta_\phi$ and $\Delta_\psi$ read
\begin{eqnarray}
\label{Deltaphi}
\Delta_\phi&=&\frac{1}{4\pi\sqrt{3}}\frac{\Lambda^2\phi_\uc}{\Mp^3}
\frac{\phi_\uc}{\phi}
\left(\frac{\phi_\uc^2}{\mu^2}+2\frac{\psi^2}{M^2}\right)^{-1}\, ,\\
\label{Deltapsi}
\Delta_\psi&=&\frac{1}{8\pi\sqrt{3}}\frac{\Lambda^2M}{\Mp^3}
\frac{M}{\psi}
\left(\frac{\psi^2}{M^2}-1+\frac{\phi^2}{\phi_\uc^2}\right)^{-1}\, .
\end{eqnarray}
These quantities are plotted in Fig.~\ref{fig:critere} in the
$(\phi,\psi)$ plane. Values such that $\Delta >1$ indicate that the
quantum effects dominate.
Let us now discuss how $\Delta_\phi$ and $\Delta_\psi$ behave in the
field plane. Clearly, $\Delta_\phi$ is infinite when $\phi=0$ for any
values of $\psi$. Therefore, the quantum effects are dominant along
that direction. In the inflationary valley $\psi=0$, which is
perpendicular to the previously mentioned direction, one has
\begin{equation}
\Delta_\phi^{\rm valley}=\frac{1}{4\pi\sqrt{3}}\frac{\Lambda^2\mu^2}{\Mp^3\phi}.
\end{equation}
This means that $\Delta_\phi>1$ as long as 
\begin{equation}
\frac{\phi}{\Mp}<\frac{\phi^{\rm valley}}{\Mp}\equiv \frac{1}{4\pi \sqrt{3}}
\frac{\Lambda ^2\mu^2}{\Mp^4}.
\end{equation}
For the parameters used in Fig.~\ref{fig:critere}, one has $\phi^{\rm
  valley}/\Mp\simeq 94$, \ie a value much larger than the upper limit
of this plot. This means that the quantum effects dominate ``very
high'' in the valley and, in particular, around the critical
point. The previous considerations explain the cross-shaped white
region centered at the origin observed in Fig.~\ref{fig:critere}. In
this regime, one expects a faithful description of the system to be
obtained only if the stochastic noises for the two coupled fields are
taken into account. In the following, we study this case, where
treating one field (for instance, the inflaton) classically and the
other (the waterfall field) stochastically is a priori not a good
approximation.

\par

\begin{figure}[t]
\begin{center}
  \includegraphics[width=8.5cm]{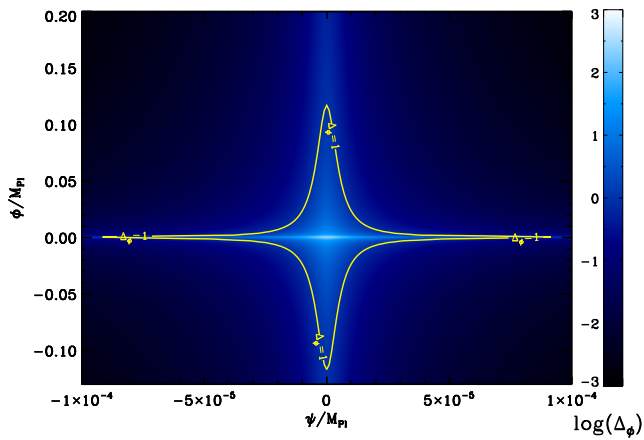}
  \caption{$\Delta_\phi$ plotted in the $\left(\phi,\psi\right)$ plane
    for the parameters $\mu=3190.4\,\Mp$, $M=\phi_\uc=0.1503\,\Mp$,
    $\Lambda=0.0005\Mp\lesssim \Lambda_\phi$. For this value of
    $\Lambda$, $\Delta_\phi$ remains small along the valley and does
    not encompass the critical point. It becomes larger than $1$ only
    in the vicinity of the origin.}
\label{fig:stochaphiSmallLambda}
\end{center}
\end{figure}

Of course, these results also depend on the parameters, in particular,
on $\Lambda$. It is interesting to determine the value of
$\Lambda\equiv \Lambda_\phi$ such that $\phi^{\rm
  valley}=\phi_\uc$. This value indicates the limit between the regime
where it is mandatory to take into account the noise both in the
inflaton and waterfall field directions and the regime where the
waterfall field is still stochastic but where it is sufficient to
treat the inflaton classically. It is given by
\begin{equation}
\label{eq:deflambdaphi}
\Lambda_\phi^2=4\pi \sqrt{3}\frac{\phi_\uc \Mp^3}{\mu^2}
\end{equation}
For our fiducial parameters, this leads to $\Lambda_\phi\simeq 5.7
\times 10^{-4}\Mp$. Obviously, for larger values of $\mu$,
$\Lambda_\phi$ is even smaller. In
Fig.~\ref{fig:stochaphiSmallLambda}, we have represented $\Delta
_\phi$ for the same parameters, except that $\Lambda =5\times
10^{-4}\Mp\lesssim \Lambda_{\phi}$. This plot confirms that the region
$\Delta _\phi>1$ covers a much smaller area which does not encompass
the critical point. In that case $\phi$ should behave almost
classically in the valley, and we will also investigate this regime in
Sec.~\ref{subsec:efold}.

\par

Let us now describe $\Delta_\psi$. It is infinite for $\psi=0$, that
is to say, in the valley. When $\psi\neq 0$, the quantum effects are
dominant when $\phi^2/\phi_\uc^2\simeq 1-\psi^2/M^2$, \ie in the
direction $\phi\simeq \phi_\uc$ perpendicular to the valley (in the
regime where $\psi/M\ll 1$). This explains the cross-shaped white
region, this time centered at the critical point; see
Fig.~\ref{fig:critere}. This time, the previous considerations do not
depend on $\Lambda$, which means that the noise in the waterfall field
should always be taken into account in the valley and around the
critical point. Since this corresponds to a very flat region of the
potential where most of the e-folds are realized, one can already
expect the inflationary dynamics to be significantly affected by the
quantum effects.

\subsection{Obstacles to a Perturbative Approach}
\label{subsec:perturbative}

Having justified that the quantum corrections play a crucial role, the
next question is how to compute them, \ie how to solve the two
Langevin equations. It is clear that an exact analytical solution is
not available. However, as proposed in
Refs.~\cite{Martin:2005ir,Martin:2005hb}, the Langevin equation can be
solved perturbatively by considering the coarse-grained field as a
perturbation on top of the classical solution. The corresponding
formalism in the case of single field inflation was presented in
Ref.~\cite{Martin:2005ir}. However, in the present case, we are in a
two-field situation, which means that both the inflaton and the
waterfall fields must be expanded according to
\begin{eqnarray}
\phi\left(N\right)&=&\phi_\mathrm{cl}\left(N\right)
+\delta\phi_1\left(N\right)+\delta\phi_2\left(N\right)+... ,\\
\psi\left(N\right)&=&\psi_\mathrm{cl}\left(N\right)
+\delta\psi_1\left(N\right)+\delta\psi_2\left(N\right)+... ,
\end{eqnarray}
where $\phi_\mathrm{cl}$ and $\psi_\mathrm{cl}$ are the classical
values. We see that the corrections to the classical solutions are
obtained by adding successive terms of higher and higher powers in the
noise. In Ref.~\cite{Martin:2005ir}, general formulas, valid at second
order, are provided, leading to a Gaussian probability density function
for the field. The validity of this approach relies on the smallness
of the stochastic effects compared to the classical ones and,
obviously, the expansion is valid only in a limited regime; see
Ref.~\cite{Martin:2005hb}. Here, we have just seen that the quantum
effects are dominant around the critical point and, therefore, there
are already reasons to guess that a perturbative approach is not very
appropriate.

\par

Moreover, one can see that the perturbative approach is technically
impossible to carry out in a multiple field situation since even the
linearized coupled stochastic differential equations cannot be
analytically solved. Indeed, in the hybrid inflation case, at first
order in the noise, they can be written as
\begin{eqnarray} 
& &\frac{\mathrm{d}\delta\phi_1}{\mathrm{d} N}+2\delta\psi_1
\left(\frac{H_{\phi\psi}}{H}-
\frac{H_\psi H_\phi}{H^2}\right)
\nonumber \\ & & 
+2\delta\phi_1\left(\frac{H_{\phi\phi}}{H}-
\frac{H_\phi^2}{H^2}\right)=\frac{H}{2\pi}\xi_\phi\left(N\right),
\\
& &\frac{\mathrm{d}\delta\psi_1}{\mathrm{d} N}
+2\delta\phi_1\left(\frac{H_{\phi\psi}}{H}-
\frac{H_\psi H_\phi}{H^2}\right)\nonumber\\ & &
+2\delta\psi_1\left(\frac{H_{\psi\psi}}{H}-
\frac{H_\psi^2}{H^2}\right)=\frac{H}{2\pi}\xi_\psi\left(N\right)\, ,
\end{eqnarray}
where $H_{\phi}$ is the derivative of
$H=\sqrt{V}/\left(\Mp\sqrt{3}\right)$ with respect to $\phi$ and the
other notations used in this equation straightforwardly follow. The
matrix of this differential system does not commute with itself at
different times $N$ and, as a consequence, one cannot solve the
coupled perturbative problem in a simple way.

\subsection{Testing the Numerical Approach}
\label{subsec:test}

\begin{figure}
\begin{center}
\includegraphics[width=0.48\textwidth,clip=true]{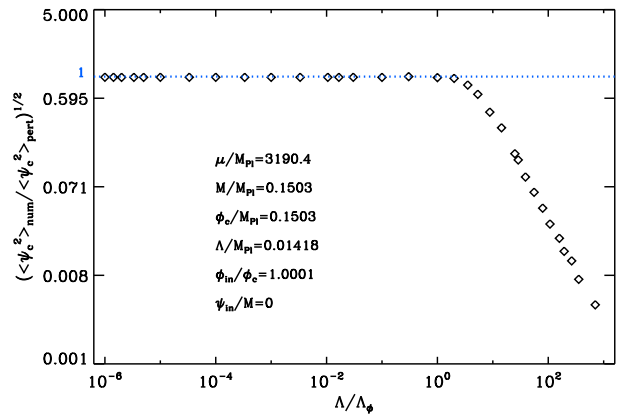}
\caption{Numerical predictions for
  $\langle\psi_{\uc}^2\rangle_{\mathrm{num}}$ normalized to
  $\langle\psi_{\uc}^2\rangle_{\mathrm{pert}}$ given by
  Eq.~(\ref{psicpert}) for different values of $\Lambda$ normalized to
  $\Lambda_\phi$ defined in Eq.~(\ref{eq:deflambdaphi}).}
\label{fig:psic}
\end{center}
\end{figure}

For all these reasons, the only method left seems to be a full
numerical integration of the stochastic inflationary equations. This
is the method used in the present article. Since the differential
equations to be solved turn out to be stiff most of the time, we use a
fourth order Rosenbrock method, monitoring a local truncation error to
adjust step sizes, that we have adapted to take into account the
presence of an extra random stochastic term. When possible, we have
also used the Euler-Muruyama method in another independent code in
order to check our numerical results. 

\begin{figure*}
\begin{center}
\includegraphics[width=0.45\textwidth,clip=true]{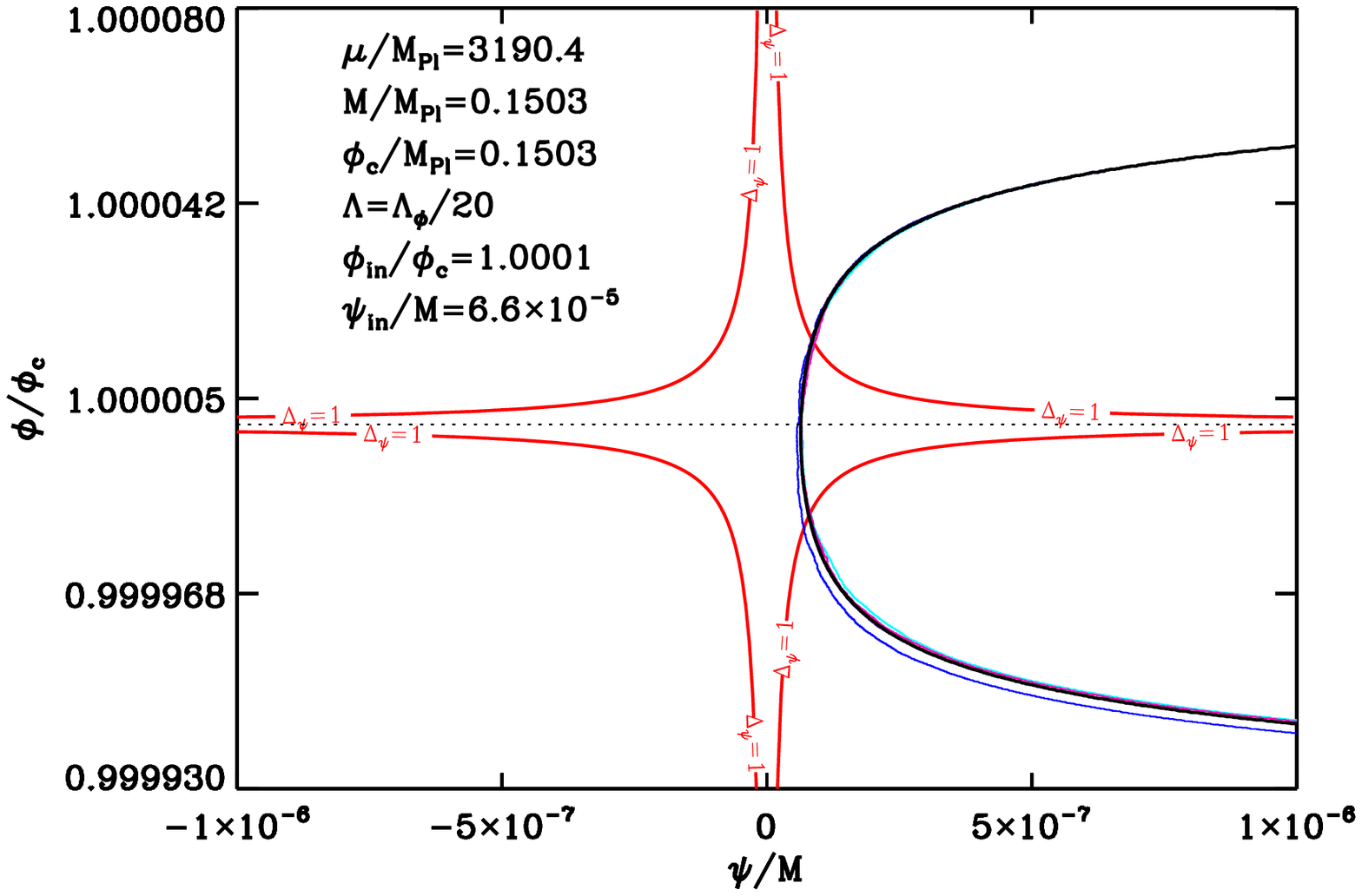}
\includegraphics[width=0.45\textwidth,clip=true]{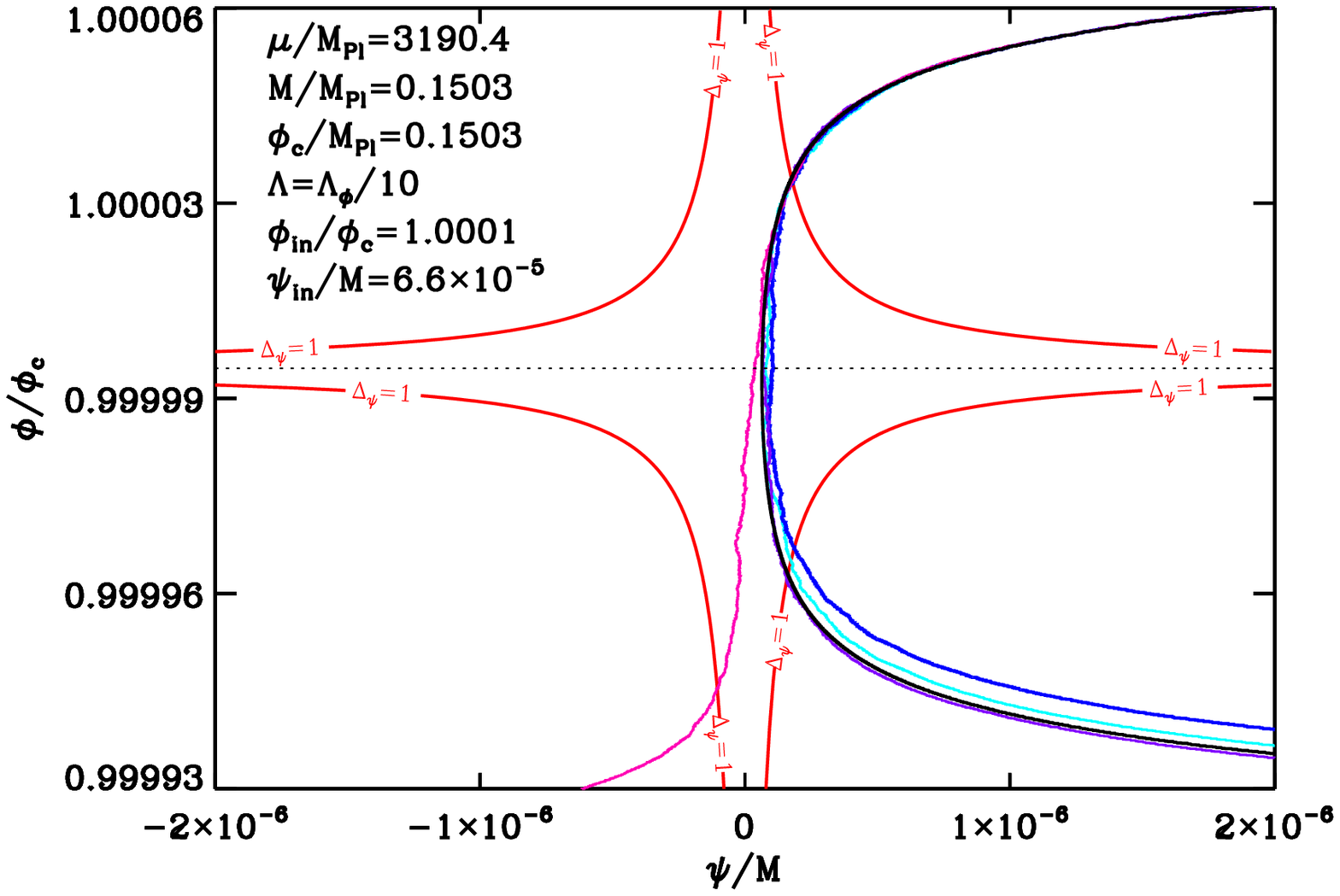}
\includegraphics[width=0.45\textwidth,clip=true]{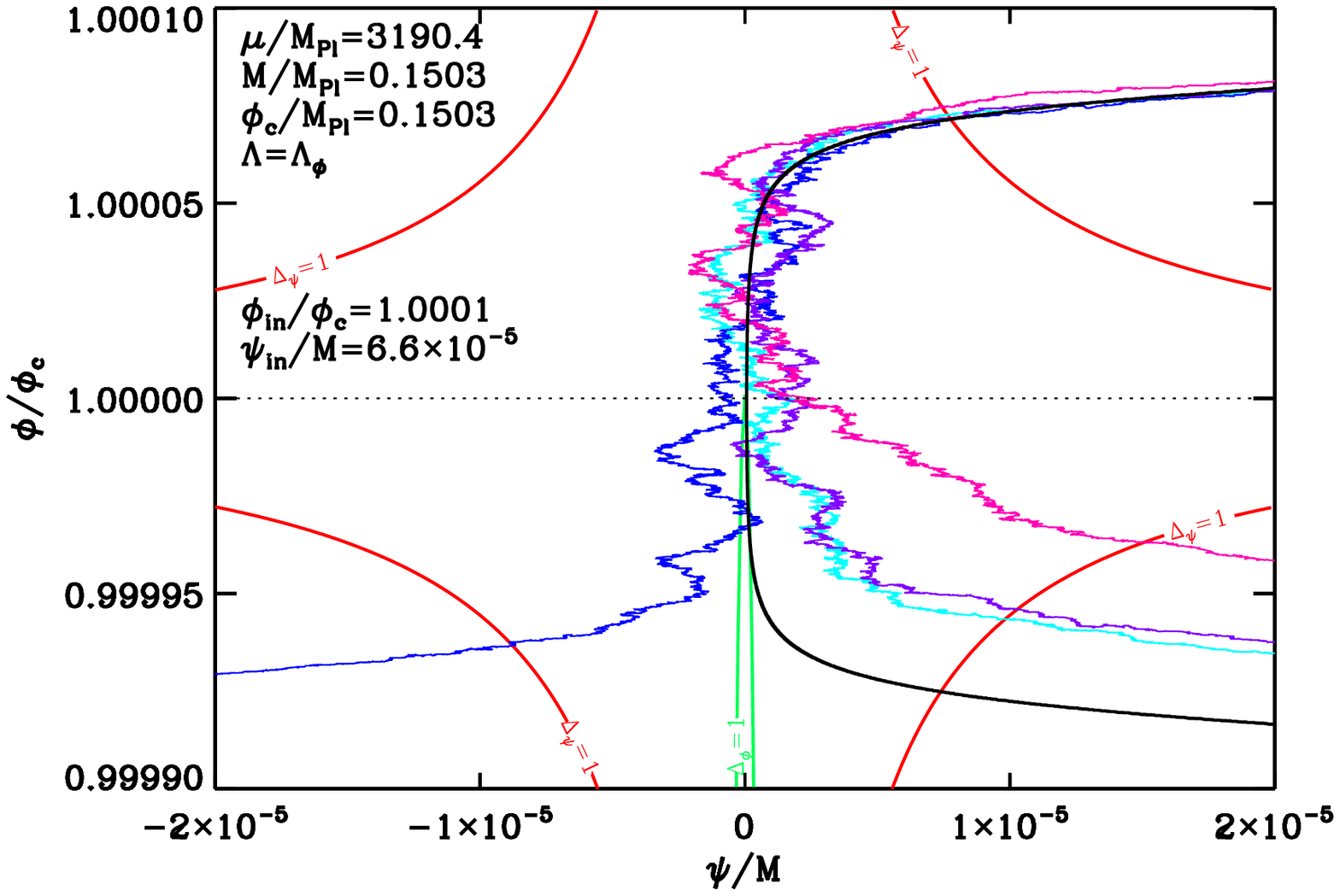}
\includegraphics[width=0.45\textwidth,clip=true]{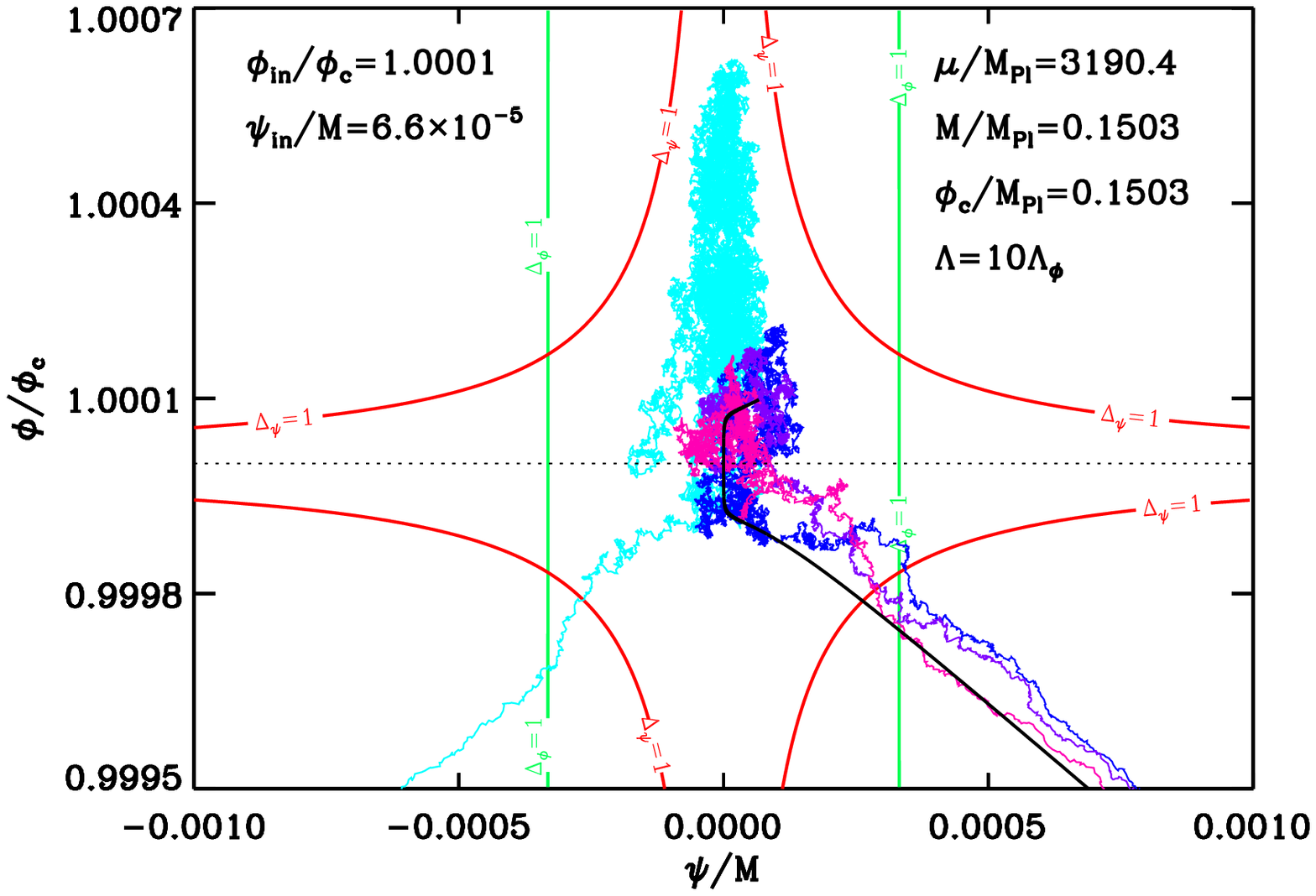}
\caption{Stochastic trajectories in field space for different values
  of $\Lambda$: $\Lambda=\Lambda_\phi/20$ (top left panel),
  $\Lambda=\Lambda_\phi/10$ (top right panel), $\Lambda=\Lambda_\phi$
  (bottom left panel), and $\Lambda=10\Lambda_\phi$ (bottom right
  panel). The solid black line represents the classical trajectory
  starting from the point $\phi_\uin=1.0001\phi_\uc$ and
  $\psi_\uin=6.6\times 10^{-5}M$. The blue, cyan, pink, and purple
  lines represent four different stochastic trajectories. The
  parameters chosen are $\mu/\Mp=3190.4$ and
  $M/\Mp=\phi_\mathrm{c}/\Mp\simeq 0.1503$. The contours
  $\Delta_\phi=1$ (solid red line) and $\Delta_\psi=1$ (solid green
  line) are also represented.}
\label{fig:trajec_stocha}
\end{center}
\end{figure*}

In this section, we describe the tests that we have performed in order
to check that our numerical codes  work properly. A first verification
of the consistency of our numerical treatment can be obtained in the
following manner. If one considers that the dynamics of $\phi$ remains
classical in the valley, then following
Refs.~\cite{GarciaBellido:1996qt,Clesse:2010iz} one can perturbatively
estimate the typical dispersion of the waterfall field distribution
(\ie in the $\psi$ direction) at the critical point. One obtains
\begin{equation}
\label{psicpert}
\langle\psi_{\uc}^2\rangle_{\mathrm{pert}}\simeq\frac{H^2\mu M}{32\pi^{3/2}\Mp^2}\, .
\end{equation}
Therefore, numerically, in the regime $\Lambda <\Lambda_\phi$ (to
ensure that the inflaton field behaves classically), one should
recover the same result. As a consequence, it is interesting to plot
the quantity
$(\langle\psi_{\uc}^2\rangle_{\mathrm{num}}/\langle\psi_{\uc}^2\rangle_{\mathrm{pert}})^{1/2}$,
where $\langle\psi_{\uc}^2\rangle_{\mathrm{num}}$ is the dispersion in
the waterfall direction (at $\phi=\phi_\uc$) obtained
numerically. This quantity as a function of $\Lambda/\Lambda_\phi$ is
represented in Fig.~\ref{fig:psic}. As it is clear from this plot,
when $\Lambda <\Lambda_\phi$, the ratio
$(\langle\psi_{\uc}^2\rangle_{\mathrm{num}}/\langle\psi_{\uc}^2\rangle_{\mathrm{pert}})^{1/2}$
is precisely $1$, thus showing that our code correctly reproduces the
known analytical result. We also see that when $\Lambda>\Lambda_\phi$,
the perturbative regime breaks down. From what we have just discussed,
the interpretation of this result is clearly that the stochastic
effects in the $\phi$ direction play a role and kick the system below
the critical point more rapidly. As a consequence, the distribution in
$\psi$ has much less e-folds to broaden than classically predicted, and
hence
$\langle\psi_{\uc}^2\rangle_{\mathrm{num}}<\langle\psi_{\uc}^2\rangle_{\mathrm{pert}}$.
This behavior is similar to what has been found in
Ref.~\cite{Enqvist:2011pt}, where it has been shown that, in case of a
multiple field inflationary dynamics with one flat direction and
several nonflat directions, the fluctuations of the nonflat
directions can be sufficient to block the growth of the
root-mean-square amplitude along the flat direction. The fact that
$\langle\psi_{\uc}^2\rangle_{\mathrm{num}}$ deviates from
$\langle\psi_{\uc}^2\rangle_{\mathrm{pert}}$ precisely at
$\Lambda=\Lambda_\phi$ is another indication of the consistency of our
numerical results.

\begin{figure*}
\begin{center}
\includegraphics[width=17cm]{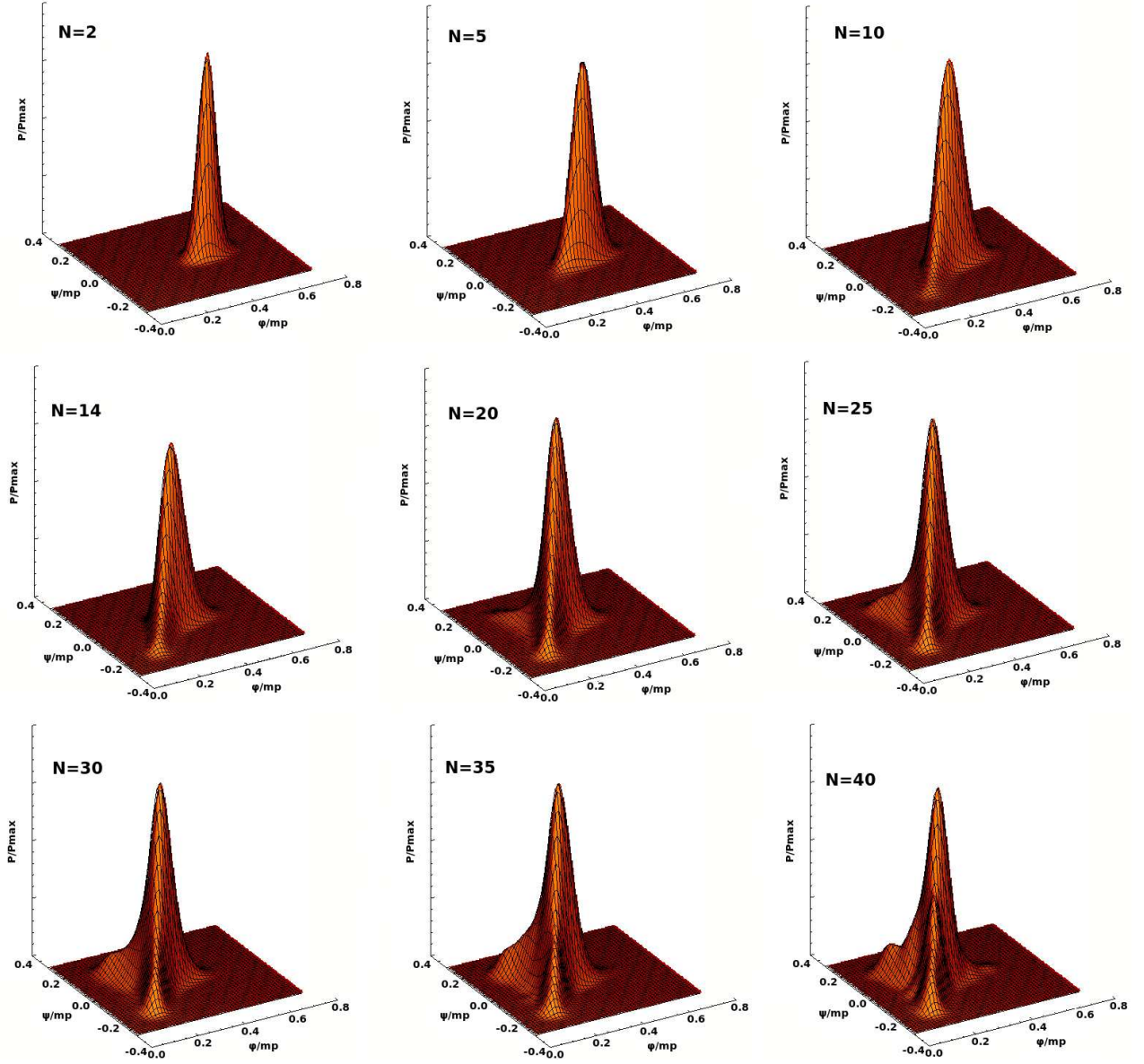}
\caption{Probability density functions in the
  $\left(\phi,\psi\right)$ plane, at different times $N$, starting
  from a Dirac distribution in the valley.  The parameters chosen are
  $\Lambda=1.06347\, \Mp$, $\phi_\uc=M=1.50398\, \Mp$, and
  $\mu=7.74597\, \Mp$.}
\label{fig:pdf2}
\end{center}
\end{figure*}

Another type of consistency check can also be performed by
investigating how given realizations behave for different values of
the parameters. We present in Fig.~\ref{fig:trajec_stocha} four
different examples, for four different values of $\Lambda$, where four
stochastic realizations (blue, cyan, purple, and pink lines) are
compared with the classical trajectory (solid black line). The top
left panel corresponds to $\Lambda=\Lambda _\phi/20$. In this case,
the noise is so small that the four stochastic realizations almost
follow the classical trajectory. The top right panel corresponds to
$\Lambda=\Lambda _\phi/10$, and the noise in the inflaton direction
still plays no role in this case (\ie the contour $\Delta_\phi=1$ does
not encompass the critical point). One sees that, when the trajectories
enter the $\Delta_\psi>1$ region, they start feeling the noise and
that one of them (the pink realization) is even expelled towards the
other global minimum. The bottom left panel corresponds to
$\Lambda=\Lambda _\phi$, and the noise is stronger as revealed by the
``shaky'' behavior of the realizations. The contour $\Delta_\phi=1$
appears but, clearly, the noise in the waterfall direction remains the
main source of stochasticity. Finally, the bottom right panel
corresponds to $\Lambda=10\Lambda _\phi$ and, this time, we are in a
regime where the noise in the two directions is \textit{a priori} important as
indicated by the contours $\Delta_\phi=1$ and $\Delta_\psi=1$. This is
especially clear for the cyan realization which climbs the
inflationary valley. Therefore, it seems fair to say that our
numerical code gives results that are completely compatible with
elementary expectations, which is an indication that it correctly
calculates the behavior of the system.

\par

Finally, by simulating a high number of realizations, we have been able
to calculate the correlation functions of various quantities of
interest as well as their probability distributions. As an example,
Fig.~\ref{fig:pdf2} shows the probability density function in the
$\left(\phi,\psi\right)$ plane, at different times $N$, starting from
a peaked distribution in the valley.  One can see that the
distribution is first roughly Gaussian and goes down the valley before
setting over the critical point with ``excrescences'' growing towards
the two minimums of the potential, rendering the distribution highly
non-Gaussian. Again, this plot confirms the previous discussion and
shows that the numerical codes used in this article are able to
reproduce expected results in regimes where it is possible to guess
(or to approximately calculate) the behavior of the system.

\par

In the following subsections, we present in more detail our numerical
results.
 
\subsection{Number of e-folds}
\label{subsec:efold}

\begin{figure}
\begin{center}
\includegraphics[width=0.47\textwidth,clip=true]{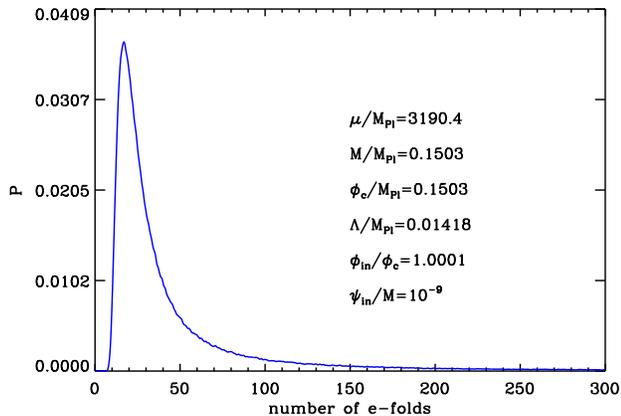}
\caption{Distribution of the total number of e-folds realized during
  inflation. Classically, $505$ e-folds are realized in the valley and
  $747$ e-folds are realized during the waterfall stage, accounting for
  a total of $1252$ e-folds. Clearly the mean value $\langle N\rangle
  \sim 50$ is very different from the classical value which
  illustrates well how important the stochastic effects in the
  vicinity of the critical point are. Despite this fact, it is also
  interesting to notice that, in the tail of the distribution, one can
  find realizations with a total number of e-folds larger than the
  classical value.}
\label{fig:NefDis}
\end{center}
\end{figure}

A first relevant well-defined physical quantity to study is the total
number of e-folds realized during inflation since it provides a
straightforward way to investigate the deviations from the classical
picture. Of course, in order to calculate this quantity, one has to
choose some initial conditions. Here, we take
$\phi_\ini/\phi_\mathrm{c}=1.0001$ and $\psi_\ini/M=10^{-9}$. The
parameters describing the shape of the potential are $\mu/\Mp=3190.4$,
$M=\phi_\mathrm{c}=0.1503\Mp$, and $\Lambda/\Mp=0.01418$. We see that
this implies $\mu M/\Mp^2>1$ and, therefore, we already know from the
previous section that the number of e-folds during the waterfall phase
will be large. As a matter of fact, it can be easily estimated upon
using Eq.~(\ref{eq:efoldswater}). The above described choice is made
in order to illustrate our point in the clearest way. It is important
to stress that choosing other initial conditions would not drastically
modify our conclusions. The classical prediction can be calculated in
the slow-roll approximation using the formulas derived above, or using
a numerical integration of the exact equations. It leads to a
trajectory such that $\sim 505$ e-folds are realized in the valley and
$\sim 747$ during the waterfall regime. The total number of e-folds is
therefore $\sim 1252$.

\begin{figure}
\begin{center}
\includegraphics[width=0.47\textwidth,clip=true]{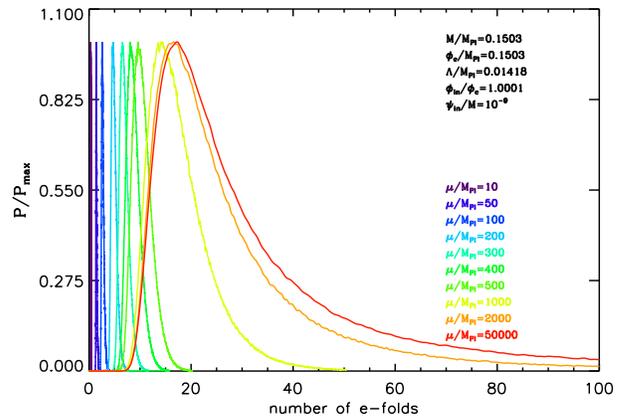}
\caption{Distribution of the total number of e-folds realized during
  inflation, normalized by its maximum value, for different values of
  $\mu$. The dependence on $\mu$ of $\langle N\rangle $ is consistent
  with the qualitative predictions of Sec.~\ref{sec:class}. One
  also notices that the dispersion of the distribution increases with
  the number of e-folds realized in the flat region. This is due to
  the fact that the quantum effects broaden the distribution.}
\label{fig:NefDisMulti}
\end{center}
\end{figure}

Then, we have computed the same quantity (for the same values of the
initial conditions and of the parameters) in the stochastic
case. Obviously, for each realization one gets a different number, and
the corresponding distribution is displayed in Fig.~\ref{fig:NefDis}.
Let us now discuss this figure. Probably, the most striking property
of Fig.~\ref{fig:NefDis} is that the distribution is peaked at a value
which is completely different from the classical prediction. This
clearly means that strong non-perturbative effects are at play. This
also emphasizes the necessity of using a full numerical approach.
Moreover, one sees that the stochastic contribution tends to diminish
the total number of e-folds.  This fact can be intuitively understood by
noticing that most e-folds are realized in the region where the
potential is very flat, around the critical point. Since this is
precisely where the stochastic terms are dominant, the quantum kicks
remove the system away from this region much faster than the classical
roll; hence a lower number of e-folds is realized in this region.

\par

The tendency to escape faster from a region where the potential is
very flat can be understood analytically on the example of small field
inflation. In this single field model, the potential is given by
\begin{equation}
V\left(\psi\right)=M^4\left[1-\left(\frac{\psi}{\mu}\right)^p\right]\, ,
\end{equation}
where $\mu$ is a mass scale. Inflation proceeds from small to large
values of the field. At the beginning of inflation, the potential is
very flat and a large number of e-folds can be realized. For $p=2$,
the slow-roll and the perturbative Langevin equations can be
integrated and solved exactly; see
Ref.~\cite{Martin:2005ir}. Following Ref.~\cite{Gratton:2005bi}, one
can then calculate the mean value of the total number of e-folds,
\begin{equation}
\langle N\rangle =-\frac{1}{2\Mp^2}\int _{\psi_\uin}^{\psi }{\rm d}\psi
\frac{\langle H\rangle}{H'_\cl}\, .
\end{equation}
In the present case and since $\phi\ll \mu $, one obtains
\begin{eqnarray}
\langle N\rangle &\simeq & N_\mathrm{class}-\frac{1}{192\pi^2}\left(\frac{M}{\Mp}\right)^4
\left(\frac{\mu}{\Mp}\right)^2\nonumber\\ & \times&
\left[\ln\left(\frac{\psi_\uin}{\psi }\right)
+\frac12\left(\frac{\psi ^2}{\psi_\uin^2}-1\right)
 \right],
\end{eqnarray} 
where $N_\mathrm{class}$ is the number of e-folds classically
realized. From the above expression, it is clear that $\langle
N\rangle$ is smaller than $N_\mathrm{class}$ since $\psi_\uin
<\psi_\uend$ in this model. This result confirms the previous
considerations: when the potential is very flat, the quantum kicks
undergone by the inflaton field push it out of the flat region and,
as a consequence, the total number of e-folds becomes smaller.

\begin{figure}
\begin{center}
\includegraphics[width=0.47\textwidth,clip=true]{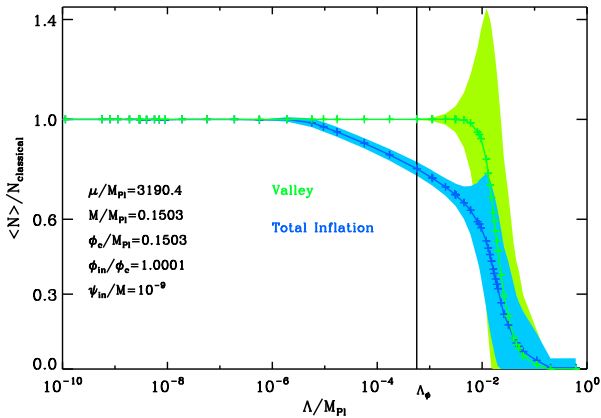}
\caption{Total number of e-folds (blue) and number of e-folds realized
  in the valley (green) as a function of $\Lambda/\Mp$ for the
  parameters indicated on the plot. The two numbers of e-folds are
  normalized to their classical counterparts. The solid lines
  represent the mean values of the distributions, while the colored
  surfaces represent the plus or minus $1$ standard deviation
  areas. The vertical solid black line indicates the value of $\Lambda
  _\phi$.}
\label{fig:NefLambda}
\end{center}
\end{figure}

Finally, it is also interesting to study how our results depend on the
parameters of the model, especially on $\mu$ and $\Lambda$. As an
example, Fig.~\ref{fig:NefDisMulti} shows the normalized distribution
of the total number of e-folds for different values of the parameter
$\mu$. From Eqs.~(\ref{Nc}) and~(\ref{eq:efoldswater}), one can see
that the classical number of e-folds realized during both the
inflationary valley and the waterfall regime increases with $\mu$,
which is consistent with the behavior of the mean values of the
stochastic distributions observed in
Fig.~\ref{fig:NefDisMulti}. Moreover, the longer the field system
stays in the flat region close to the critical point, the more its
distribution gets stochastically broadened. This means that the
dispersion of the distribution should evolve in a similar manner,
which is exactly what is seen in Fig.~\ref{fig:NefDisMulti}. Moreover,
if one keeps increasing $\mu$, one observes that the mean value of the
distribution saturates at a value of $\simeq 60$ e-folds.

\begin{figure}
\begin{center}
\includegraphics[width=0.47\textwidth,clip=true]{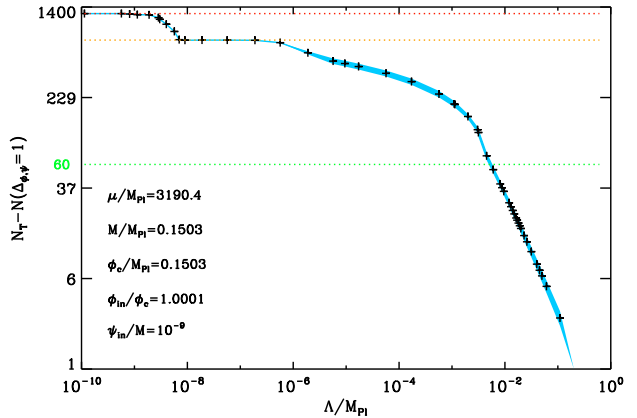}
\caption{Number of e-folds realized between the exit of the stochastic
  regime (\ie the moment where $\Delta_\phi$ and $\Delta_\psi$ are
  both smaller than $1$) and the end of inflation. The cruxes stand
  for the mean values of the distributions, and the colored surfaces
  stand for the plus or minus $1$ standard deviation areas. The
  horizontal dotted red line represents the total number of e-folds
  calculated in the absence of noise. The horizontal dotted orange line is
  the number of e-folds realized in the waterfall region in the absence of
  noise.  Finally, the horizontal green dotted line represents the
  minimal number of e-folds required for inflation to be successful,
  \ie $60$.}
\label{fig:NefLambdaClassEnd}
\end{center}
\end{figure}

As mentioned above, we have also studied how our results depend on
$\Lambda$. In the vacuum dominated regime, $\Lambda $ is directly
related to the energy scale of inflation. This quantity is constrained
by the big bang nucleosynthesis and by the observations of the
CMBR, namely,
\begin{equation}
  10^{-17}\Mp\lesssim \Lambda\lesssim 5\times 10^{-2}\Mp\, .
\end{equation}
The previous figures correspond to a regime where $\Lambda>\Lambda
_\phi$ (recall that, for the values of the parameters chosen here, we
have $\Lambda_\phi\simeq 5.7\times 10^{-4}\Mp$) since we wanted to study
a regime where the noise in the two-field directions is
important. However, one can also wonder if the previous conclusions,
especially the fact that $\langle N\rangle <N_{\rm class}$, still hold
in the regime $\Lambda<\Lambda_\phi$. In Fig.~\ref{fig:NefLambda}, we
have computed the total number of e-folds as a function of $\Lambda$.
The first thing we notice in this plot is that, for $\Lambda \gtrsim
\Lambda_\phi$ (indicated by the vertical black line), the total number
of e-folds and the number of e-folds in the valley are smaller than
their classical counterparts. This is of course compatible with the
previous considerations. In the regime $10^{-6}\Mp\lesssim \Lambda
\lesssim \Lambda_\phi$, the number of e-folds in the valley is equal
to its classical counterpart as expected since the inflaton behaves
classically but the total number of e-folds, hence the number of
e-folds in the waterfall regime, is still smaller than $N_{\rm
  class}$. So even in the absence of noise along the inflaton direction,
the conclusion obtained before remains valid. Finally, for $\Lambda
\lesssim 10^{-6}\Mp$, the noise is so small that the stochastic and
classical number of e-folds are equal.

\begin{figure}
\begin{center}
\includegraphics[width=0.47\textwidth,clip=true]{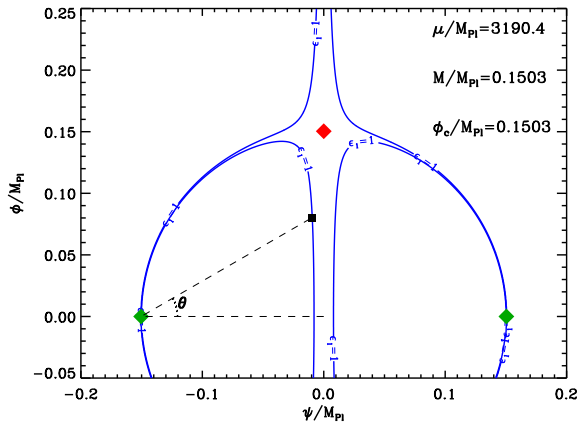}
\caption{Level lines $\epsilon_1=1$ in the $\left(\phi,\psi\right)$
  plane. The parameters chosen are the same as in the other plots. The
  red rhombus sits at the critical point, while the two green ones are
  located at the two minima. The definition of the angle $\theta$ is
  shown for a trajectory leaving the inflationary region (\ie crossing
  the $\epsilon_1=1$ line) at the black square.}
\label{fig:epsilon1}
\end{center}
\end{figure}

In Fig.~\ref{fig:NefDis}, we have seen that, for $\Lambda=0.01418\Mp$,
one has $\langle N\rangle \simeq 50$. Since the scales of
astrophysical interest today left the Hubble radius during inflation
about $50-60$ e-folds before the end of inflation, this would mean
that we could have a direct observational window on the stochastic
regime. In fact, this is not so because $\Delta =1$ means that
$H^2/\epsilon_1\simeq 1$. But $H^2/\epsilon_1$ is precisely the
overall normalization of the density perturbations power spectrum
which is observed to be $\simeq 10^{-5}$. So, in fact, this shows that
the value $\Lambda=0.01418\Mp$ is simply excluded by the CMBR
measurements. For this reason, it is interesting to plot the number of
e-folds performed between the moment the system becomes classical (\ie
when $\Delta _\phi$ and $\Delta_\psi$ are both smaller than $1$) and
the end of inflation. If this number is smaller than $\sim 50-60$,
this means that the corresponding value of $\Lambda$ is excluded due
to the above argument. The plot is represented in
Fig.~\ref{fig:NefLambdaClassEnd}. We see that for $\Lambda\gtrsim
10^{-2}\Mp$, the number mentioned above is indeed smaller than
$60$. All these values are therefore excluded. This means
that for values of $\Lambda$ such that
$\Lambda_\phi\lesssim\Lambda\lesssim10^{-2}\Mp$, we are in a regime
where the perturbations are not too large to be directly in
contradiction with the CMBR (of course this does not guarantee that
the correct normalization can be obtained) and where it is mandatory
to take into account the stochastic effects in the two-field
directions. If $10^{-6}\Mp <\Lambda<\Lambda_{\phi}$, the stochastic effects
dominate only in the $\psi$ direction, both in the valley and the waterfall phase.
For $10^{-8}\Mp\lesssim
\Lambda\lesssim 10^{-6}\Mp$, the waterfall regime becomes completely
classical and, finally, for $\Lambda\lesssim 10^{-8}\Mp$, the noise
becomes so small that the full evolution in the valley and in the
waterfall region can be described classically.

\par

Let us end this subsection with some remarks. We have seen that the
stochastic number of e-folds is smaller than its classical counterpart
as soon as $\Lambda \gtrsim 10^{-6}\Mp$. If $\Lambda \gtrsim
\Lambda_\phi\simeq 5.7\times 10^{-4}\Mp$, we are in a two-field
regime. Moreover, if $\Lambda \gtrsim 10^{-2}\Mp$, the stochastic
effects are so strong that the model is in contradiction with the
amplitude of the CMBR fluctuations. We expect these conclusions to be
very roughly independent of the choice of the other parameters,
provided, of course, that one remains in the regime described in
Sec.~\ref{subsec:priors}. In fact, to go further, one should explore
the full parameter space, and one should carefully apply Cosmic Background Explorer (COBE) normalization to the
model. When the waterfall regime plays an important role, this is not
a trivial task.

\subsection{Inflation Exit Point}
\label{subsec:exit}

\begin{figure}
\begin{center}
\includegraphics[width=0.47\textwidth,clip=true]{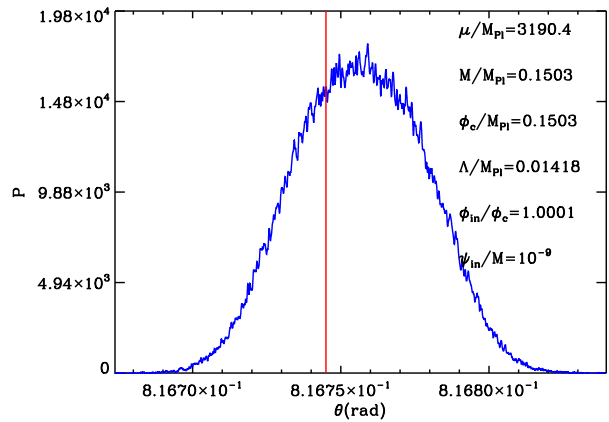}
\caption{Numerical distribution of the exit angle $\theta$. The
  parameters are the same as in the other figures. The red line
  corresponds to the classical prediction. We notice that the
  distribution is extremely peaked, $\Delta\theta/\theta\sim10^{-3}$.}
\label{fig:ThetaDis}
\end{center}
\end{figure}

\begin{figure}
\begin{center}
\includegraphics[width=0.47\textwidth,clip=true]{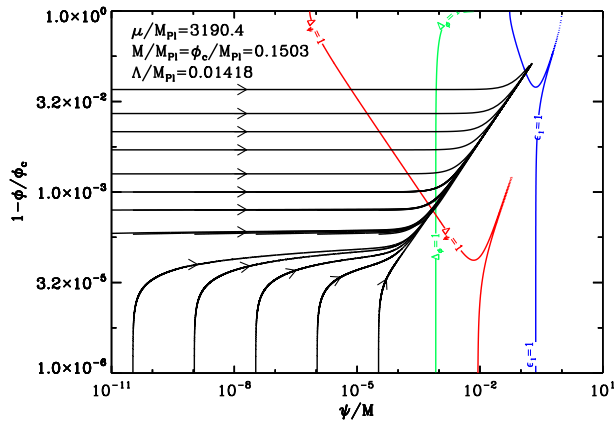}
\caption{Flow map of the classical slow-roll dynamics equations.
  Regardless of their initial conditions, all the trajectories end up
  at the same exit point.}
\label{fig:classicalflow}
\end{center}
\end{figure}

Another relevant physical quantity is the inflation exit point, \ie
the location in the field space where inflation stops. The details of
the subsequent (p)reheating phase strongly depend on these initial
conditions, which are therefore important physical
quantities~\cite{GarciaBellido:1997wm,Finelli:2000ya}. Inflation stops
when the system crosses the $\epsilon_1=1$ level line in the field
space. As a consequence, the exit point is necessarily located on this
level line. It can be characterized by the angle $\theta$ between the
line joining the closest minimum to the origin and the line joining
this same minimum to the exit point. This parametrization and the
$\epsilon_1=1$ contours are represented in
Fig.~\ref{fig:epsilon1}. Since the stochastic effects taking place in
the valley quickly render the distribution symmetrical in $\psi$, the
two minima are in fact put on an equal footing. Using the same method
as before, one can calculate the classical prediction for this angle
$\theta$ and compare it with the corresponding stochastic
distribution. The result is shown in Fig.~\ref{fig:ThetaDis}.

\par 

Here again, several comments are in order. First, unlike the
distribution of the number of e-folds, the classical prediction lies
within the stochastic distribution. It is even more remarkable that
the distribution is very narrow. \textit{A priori}, this is a surprising fact
since the stochastic realizations in the vicinity of the critical
point are extremely noisy, as can be seen in
Fig.~\ref{fig:trajec_stocha}. Even if the stochastic trajectories are
very different from one realization to another and spread over a large
area in the field space, they eventually gather at the end of the
waterfall phase to exit inflation at nearly the same point. We
interpret this property in the following manner. Around the critical
point, as was already mentioned before, the noise is quite strong and
it quickly kicks the fields out of this region. As a consequence, the
fields eventually land in a region where the noise is
subdominant. Therefore, from that point, the fields will follow a
classical trajectory. If there is a classical attractor, all the
trajectories will converge towards this particular path, and inflation
will always stop at the same point. This analysis is confirmed by
Fig.~\ref{fig:classicalflow}, where we have plotted in the field plane
the flow lines of the classical equations of motion, Eqs.~(\ref{KGphi})
and~(\ref{KGpsi}). As can be seen in the figure, after crossing the
$\Delta_\psi=1$ and $\Delta_\phi=1$ level lines, which implies that the
fields enter a region where the stochastic terms are subdominant, all
the classical trajectories merge into a single one before crossing the
$\epsilon_1=1$ level line and thus exiting the inflationary
region. Then, one can check that this point corresponds to the angle
$\theta$ singled out in Fig.~\ref{fig:ThetaDis}.  This classical
attractiveness can also be formally established by studying the Lyapounov
exponent, in the direction orthogonal to the flow tangents.  We
conclude that, despite the strong quantum effects undergone by the
fields during the inflationary phase, the exit point is always the
same (approximately, of course) in hybrid inflation. Moreover, this
point turns out to be the classical one which provides a
straightforward way to calculate its location.

\section{Conclusion}
\label{sec:conclusion}

Let us now summarize our main findings. We have found that the quantum
effects play an important role in hybrid inflation, especially in the
vicinity of the critical point (this seems to be a general feature of
multiple field models of inflation, as soon as flat directions are present
in the potential). As a consequence, the classical
picture presented in Sec.~\ref{sec:class} has to be substantially
modified. This can be done in the framework of stochastic inflation,
where the inflationary dynamics is driven by two coupled Langevin
equations. Given that the stochastic effects can be strong in the two
directions in field space, we have used a numerical approach to solve
these equations. Then, we have derived the distributions of two
relevant quantities, namely the total number of e-folds realized
during inflation and the exit point. We have shown that, when the
stochastic noise plays a role, the distribution of the number of
e-folds is peaked at a value which is different from the classical
prediction. This is due to the fact that, in the neighborhood of the
critical point, the potential is very flat and the quantum kicks
quickly move the system away from this region. On the other hand,
the distribution of the exit point of inflation leads to conclusions
which are apparently at odds with this picture since it is extremely
peaked over the classical prediction. But, in fact, this property is
due to the attractiveness of the classical flow and is not at all in
contradiction with the previous considerations.

\par

An important question that remains to be addressed in more detail is
the impact of our results on the observable predictions of hybrid
inflation. For instance, it would be of utmost importance to study how
the quantum effects can modify the power spectra of cosmological
fluctuations. It was recently emphasized in Ref.~\cite{Clesse:2008pf}
that hybrid inflation can lead to a red spectrum, and it would be
interesting to investigate the influence of the quantum effects on
this prediction, both for the adiabatic and entropy
modes~\cite{Levasseur:2010rk}. Also, the explicit computation of the
probability density functions in the field space provides us with a
means to calculate the non-Gaussianities of this model. We intend to
come back to those issues in the future.  Maybe the most important
conclusion of our work is that the richness of multiple inflation -
namely, the presence of entropy modes a priori produced during the
waterfall regime, the highly non-trivial phase of preheating, the
strong quantum effects - implies that it is not simple to derive the
corresponding observable predictions and that, most of the time, these
ones cannot be obtained in a simple single field effective model. This
is an important conclusion that one should keep in mind when analyzing
the future high accuracy CMB data and their implications for
inflation.

\acknowledgements

We would like to thank S.~Clesse and C.~Ringeval for useful discussions.

\bibliography{biblio}

\end{document}